\documentclass[showpacs,amsmath,amssymb,showkeys,pre]{revtex4}
\usepackage{amsmath}
\usepackage{amssymb}
\usepackage{epsfig}

\newcommand{\e}{\epsilon}
\newcommand{\E}{\mathbb{E}}
\renewcommand{\P}{\mathcal{P}}
\newcommand{\Q}{\mathcal{Q}}
\newcommand{\N}{\mathcal{N}}

\begin{document}

\title{Random multi-index matching problems}

\author{O. C. Martin, M. M\'ezard and O. Rivoire}

\affiliation{Laboratoire de Physique Th\'eorique et
Mod\`eles
Statistiques \\
Universit\'e\ Paris-Sud, B\^at. 100, 91405 Orsay
Cedex, France}

\date{July, 2005}

\begin{abstract}
The multi-index matching problem (MIMP) generalizes the well known matching
problem by going from pairs to $d$-uplets. We use the cavity method from
statistical physics to analyze its properties when the costs of the $d$-uplets
are random. At low temperatures we find for $d\geq 3$ a frozen glassy phase
with vanishing entropy. We also investigate some properties of small samples
by enumerating the lowest cost matchings to compare with our theoretical
predictions.
\end{abstract}



\maketitle

\section{Introduction}

The statistical properties of random combinatorial optimization problems can
be studied from a number of angles, with tools depending on the discipline.
Recent years have however witnessed a convergence of interests and techniques
across mathematics, computer science and statistical physics. An archetype
example is the matching problem with random edge weights, defined as follows:
suppose one has $M$ different jobs and $M$ people to perform them, one person
per job, and let $c_{ij}$ be the cost when job $i$ is executed by person $j$;
the 2-index matching problem consists in assigning jobs to people in such a
way as to minimize the total cost. The statistical properties of the optimal
matching when the cost $c_{ij}$ are drawn independently from a common
distribution were found two decades ago using the
replica~\cite{MezardParisi85} and the cavity~\cite{MezardParisi86} methods.
These two non-rigorous statistical physics approaches have recently been used
to tackle a number of computationally more difficult problems such as
satisfiability or graph coloring, but the 2-index matching problem sets apart
for belonging to one of the very few problems where such predictions have been
rigorously confirmed~\cite{Aldous01}.

In this work, we take the statistical physics approach and study the
properties of a generalization of the 2-index to multi-index matching problems
(MIMPs) where the elementary costs are now associated with $d$-uplets,
representing for example persons, jobs and machines when $d=3$. At variance
with the 2-index matching, $d$-index matching problems with $d\geq 3$ are
NP-hard. We show here that their low lying configurations also have a
different, glassy, structure whose description requires replica symmetry to be
broken. Remarkably, the replica symmetry breaking scheme differs from the
common picture that has emerged from the study of other optimization problems
such as the coloring \cite{MuletPagnani02} and satisfiability problems
\cite{MezardZecchina02}. In particular, a na\"{\i}ve application of the 1-RSB
cavity method at zero temperature \cite{MezardParisi02}, which successfully
solves these two problems, is here doomed to fail. The reason for this will be
traced back to the presence of ``hard constraints''. By unraveling this
specificity, we put forward arguments whose relevance goes beyond matching
problems; they indicate when a similar scenario can be expected on other
constrained systems. The particularly simple glassy structure that we find is
also of interest from the interdisciplinary point of view: in conjunction with
the rigorous formalism available for the 2-index case, it places MIMPs in a
choice place for working out a most awaited mathematical understanding of
replica symmetry breaking.

The present paper provides an extensive account of our results on the MIMPs,
some of which have already been mentioned in~\cite{MartinMezard04}. The paper
is organized as follows. We first define precisely multi-index matching
problems, and briefly review the past approaches from physics, mathematics and
computer science that were developed mainly to address the 2-index case. Then
we start our statistical study by establishing the scaling of the minimal cost
as a function of the number of variables and by providing a lower bound from
an annealed calculation. A large part of the paper is then devoted to present
our implementation of the cavity method to matching problems, including a
detailed discussion of its relations with the rigorous formalism proposed by
Aldous; we explain why and how replica symmetry must be broken when $d\geq 3$,
in order to account for the presence of a frozen glassy phase. Finally, the
last section is dedicated to a numerical analysis of small samples that
provides support to the proposed scenario.

\section{Multi-index matchings}

\subsection{Definitions}

\begin{figure}
   \begin{minipage}[c]{.46\linewidth}
   \centering
\epsfig{file=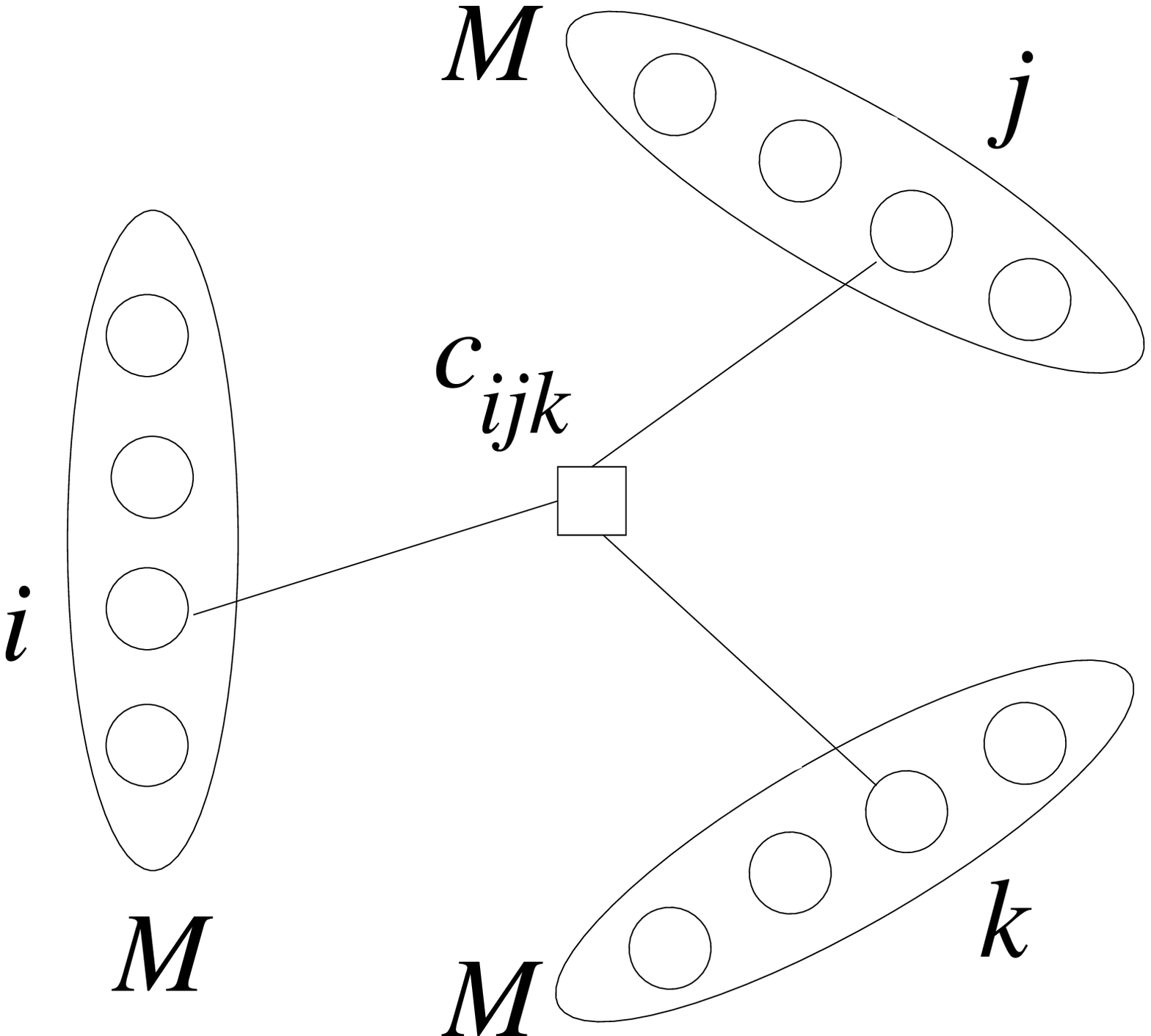,height=4cm} \caption{\small
{\it Tripartite matching problem.} Factor graph
representation :
the hyperedges, or factor nodes, are represented with
squares.}\label{fig:3partite}
   \end{minipage} \hfill
   \begin{minipage}[c]{.46\linewidth}
   \centering
\epsfig{file=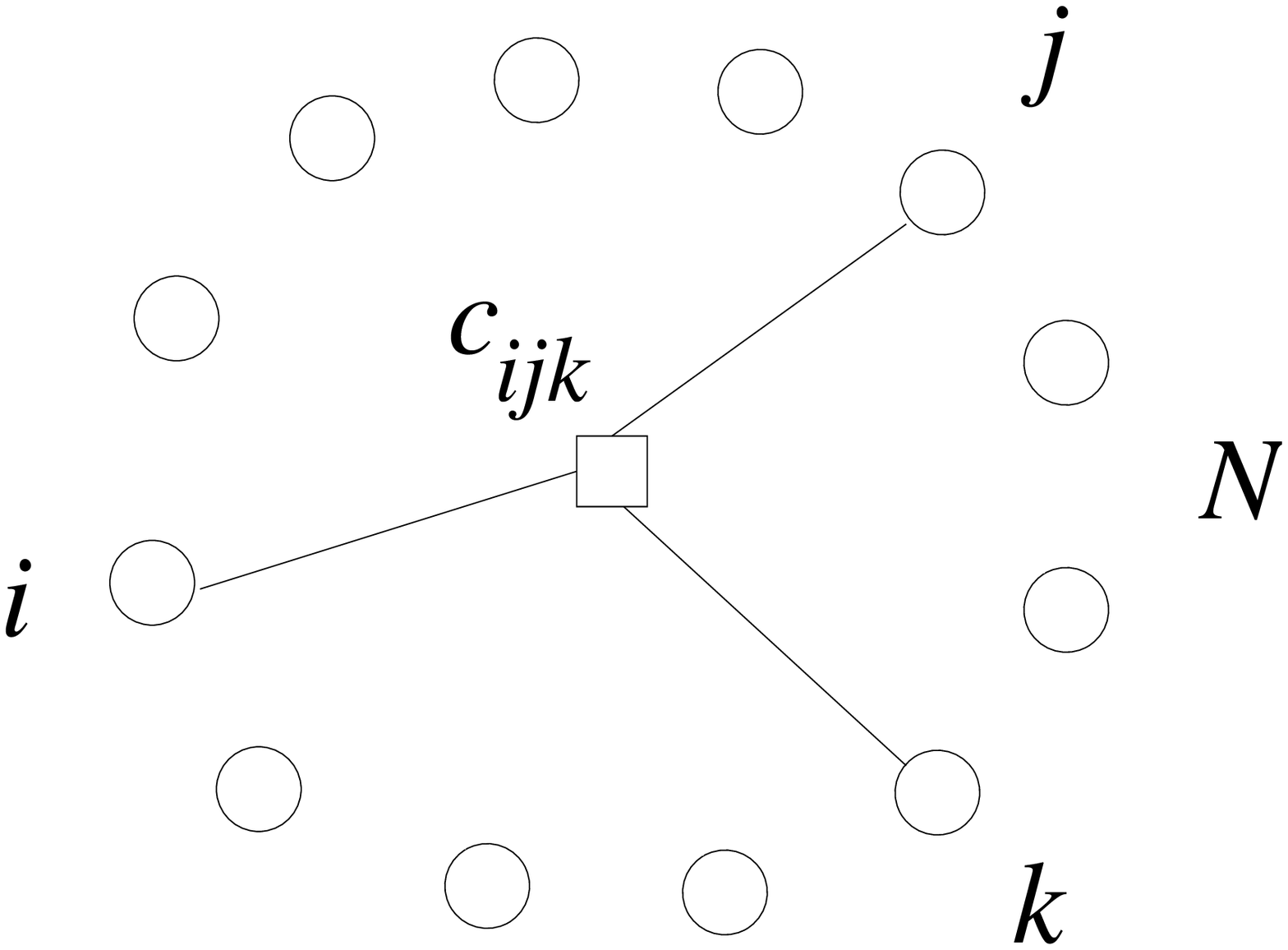,height=4cm} \caption{\small
{\it Simple 3-index matching problem.} The factor
graph representation is similar to
the tripartite case.}\label{fig:3simple}
   \end{minipage}
\end{figure}

Two classes of MIMPs can be distinguished, $d$-partite matching problems and
simple $d$-matching problems, whose asymptotic properties will be shown to be
related. We first start with the {\it $d$-partite matching problem} that
corresponds to the version alluded to in the introduction. An instance
consists of $d$ sets, $A_1$,\dots, $A_d$, of $M$ {\it nodes} each, and a cost
$c_a$ is associated with every $d$-uplet $a=\{i_1,\dots,i_d\}\in
A_1\times\dots\times A_d$. Graphically, it is represented by a factor graph as
shown in Fig.~\ref{fig:3partite} with hyperedges (factor nodes) joining
exactly one node from each ensemble. A matching $\mathcal{M}$ is a maximal set
of disjoint hyperedges, such that each node is associated to one and only one
hyperedge of the matching; it can be described by introducing an occupation
number $n_a\in\{0,1\}$ on each hyperedge $a$, with the correspondence
\begin{equation}
a\in\mathcal{M}\Leftrightarrow n_a=1.
\end{equation}
The condition for a set of hyperedges to be a matching
can then be written
\begin{equation}\label{eq:constraintdpart}
\forall r= 1, \dots,d, \quad \forall i_r \in A_r,\quad
\sum_{a\ : \ i_r\in a}n_a=1.
\end{equation}

The $d$-partite matching problem consists in finding
the matching with minimal total cost,
\begin{equation}
C_M^{(d)}=\min_{\{n_a\}}\sum_a c_a n_a
\end{equation}
with the $\{n_a\}$ subject to the constraints (\ref{eq:constraintdpart}). We
consider here the random version of the problem, where the costs $c_a$ are
independent identically distributed random variables taken from a distribution
$\rho(c)$, and we are interested in the typical value of an optimal matching
in the $M\to\infty$ limit. For definiteness, we take for $\rho$ the uniform
distribution in $[0,1]$, but the asymptotic properties of $d$-matchings depend
only on the behavior of $\rho$ close to $c=0$, and are identical for all
distributions $\rho$ with $\rho(c)\sim 1$ as $c\to 0$, such as the exponential
distribution, $\rho(c)=e^{-c}$. The case $\rho(c)\sim c^r$, $r>0$, can be
treated along the same lines, but gives different quantitative results.

A variant of this setup is the {\it simple} $d$-matching problem, where a
unique set of $N$ nodes, with $N$ being a multiple of $d$, is considered and a
cost is associated to each $d$-uplet of nodes (see Fig.~\ref{fig:3simple}).
The $d$-partite case can be seen as a particular instance of a simple
$d$-matching problem where the hyperedges joining more than one node of any
$A_i$ are given an infinite cost. Simple $d$-matchings problems are formulated
as finding
\begin{equation}
L_N^{(d)}=\min_{\{n_a\}}\sum_a c_a n_a
\end{equation}
under the constraints
\begin{equation}\label{eq:constraint}
\forall i= 1,\dots, N, \quad \sum_{a\ :\ i\in a}n_a=1.
\end{equation}

Before presenting our analysis of random matching problems by means of an
adaptation of the cavity method for finite connectivity statistical physics
models, we briefly review past approaches to the subject, with an emphasis on
open questions that motivated the present study.

\subsection{Physical approach}

The 2-index matching problem was the first combinatorial optimization problem
to be tackled with the replica method, an analytical method initially
developed in the context of spin glasses \cite{MezardParisi87b}. In the paper
\cite{MezardParisi85}, M\'ezard and Parisi analyzed both the simple and
bipartite matching problems for cost distributions $\rho$ with $\rho(c)\sim
c^r$ as $c\to 0$. Using replica theory within a replica symmetric Ansatz, they
derived the minimal total cost; thus, for the bipartite matching with $r=0$,
they predicted $\lim_{M\to\infty}\mathcal{C}^{(2)}_M=\pi^2/6$; moreover, they
obtained the distribution of cost in the optimal matching. Support in favor of
their prediction has first come from numerical results and from an analytical
study of the stability of the replica symmetric
solution~\cite{MezardParisi87,ParisiRatieville02}. This last analysis further
yields the leading corrections of order $1/N$ for the value of the minimum
matching.

Interestingly, the same results can be reobtained using a variant of the
cavity method based on a representation of self-avoiding walks using
$m$-component spins~\cite{MezardParisi86}. This alternative formulation,
avoiding the bold prescriptions of replica theory, furthermore suggests that,
if the cost of the hyperedges connected to a given node are ordered from the
lowest to the highest, the probability for the $k$-th hyperedge to be included
in the optimal matching is $2^{-k}$~\cite{ParisiRatieville01}, as first
conjectured from a numerical study~\cite{HoudayerBoutet98}.

\subsection{Mathematical approach}

Replica theory, while a powerful tool to obtain analytical formulae, is not a
rigorously controlled method, and its predictions have only the status of
conjectures within the usual mathematical standards. For the 2-index matching
problem with $r=0$ however, the results mentioned above (value of the optimal
matching, distribution of costs, and probability of inclusion of $k$-th
hyperedge) have all been confirmed by a rigorous derivation, due to
Aldous~\cite{Aldous01}. His contribution also includes the proof an {\it
asymptotic essential uniqueness} property that mathematically expresses the
fact that replica symmetry indeed holds for this problem. The weak convergence
approach~\cite{AldousSteele03} on which the proof is built is closely related
to the cavity method we will employ, and the relations between the two
formalisms will be discussed in Sec.~\ref{sec:t=0}. Confirmation of the
$\zeta(2)=\pi^2/6$ value for the bipartite assignment problem also comes from
the recent proofs~\cite{LinussonWastlund03,NairPrabhakar03} of a more general
conjecture formulated by Parisi~\cite{Parisi98}; this conjecture states that,
for the bipartite matching with exponential distribution of the costs,
$\rho(c)=e^{-c}$, the mean optimal matching for {\it finite} $M$ is
$\sum_{k=1}^M k^{-2}$.

These mathematical contributions are part of a more ambitious program aiming
at developing rigorous proofs and possibly a rigorous framework of the replica
and cavity methods. Interestingly, Talagrand, one of the prominent advocate of
this program, devotes the last chapter of his book on the
subject~\cite{Talagrand03} to the 2-index matching problem, stressing that, in
spite of the major advances mentioned, it stays a particularly challenging
issue. Indeed, finite temperature properties have so far resisted to
mathematical investigations, even in the limit of high temperature, that has
been successfully addressed in other spin-glass like
models~\cite{Talagrand03}. We shall comment on the peculiarities of matchings
with respect to other constrained systems in Sec.~\ref{sec:frozen}. It is our
hope that our work not only provides new challenging conjectures, but also
suggests some hints for solving unanswered preexisting mathematical questions.

\subsection{Computer science approach}

If analytical studies of random $d$-matchings by statistical physicists and
mathematicians have been restricted up to now to the $d=2$ case, $d$-index
matching problems with $d>2$ have a longer history in the computer science
community. $d$-partite extensions of the bipartite matching problem were
introduced in 1968 under the name {\it of multidimensional assignment
problems}~\cite{Pierskalla68}; they are also referred in the literature as
{\it multi-index assignment problems}, and, more specifically, as multi-index
{\it axial} assignment problems (to distinguish them from the so-called {\it
planar} versions~\cite{Spieksma00,Burkard02}). MIMPs, as we call them (for
multi-index matching problems), have a number of practical applications. The
most commonly cited one is for data association in connection with
multi-target tracking~\cite{Poore94}. Besides a major interest for real-time
air traffic control, such approaches are for instance helpful for tracking
elementary particles in high energy physics
experiments~\cite{PusztaszeriRensing95}.

From the algorithmic complexity point of view, matching problems have also a
pioneering role since the 3-index matching problem was among the first 21
problems to be proved NP-complete~\cite{Karp72}. In contrast, polynomial
algorithms are known that solve 2-index matching
problems~\cite{PapadimitriouSteiglitz82}. Note that being based on a worst
case analysis, NP-hardness is however only a necessary condition for hard
typical complexity, which is the issue which interests us here. Due to their
intrinsic algorithmic difficulty and to the broad range of their applications,
generalized assignment problems are the subject of numerous studies in the
computer science community; we refer to the
reviews~\cite{Spieksma00,Burkard02} for additional information and references.

\section{Scaling and a lower bound}

The first task in studying random optimization problems is to determine the
scaling of the optimal cost with the number of
variables~\cite{VannimenusMezard84}. Here, we address this issue for the two
variants of MIMPs, the multi-partite and simple multi-index matching problems.
The scaling is inferred from an heuristic argument, and confirmed by an
annealed calculation (first moment method) yielding a lower bound. This leads
us to a statistical physics formulation that encompasses the two versions of
MIMPs.

\subsection{Scaling}

The statistical physics approach of combinatorial optimization problems
consists in defining the energy $E(\mathcal{M})$ of each admissible solution,
here a $d$-matching $\mathcal{M}$, as its total cost,
$E(\mathcal{M})=\sum_{a\in\mathcal{M}}c_a$, and in determining the minimal
total cost, identified with the ground-state energy, by looking at the zero
temperature properties of the system. For $d$-matchings, the corresponding
Hamiltonian
\begin{equation}
\mathcal{H}[\{n_a\}]=\sum_a c_a n_a
\end{equation}
defines a lattice gas model, where the particles are occupying the hyperedges.
The constraints (\ref{eq:constraintdpart}) or (\ref{eq:constraint}) implement
a hard-core interaction between the particles: two ``neighboring'' hyperedges
are not allowed to be occupied simultaneously. To have a sensible statistical
physics model, the ground state has to be extensive, {\it i.e.}, proportional
to $M$ in the $d$-partite case, and to $N$ in the simple case. We propose here
a heuristic argument to determine how $\E[C_M^{(d)}]$ and $\E[L_N^{(d)}]$
scale with $M$ and $N$ respectively, where $\E[\cdot]$ represents the average
over the different realizations of the costs. The central (local) quantity
that monitors the scaling behavior is the number of hyperedges to which a
given node belongs, noted $W_\Lambda^{(d)}$ ($\Lambda=M$ or $N$). Indeed, with
the costs uniformly distributed in [0,1], the lowest costs to which a node can
be associated are of order $1/W_\Lambda^{(d)}$ and the optimal matching is
expected to scale like $\Lambda/W_\Lambda^{(d)}$. Thus, for $d$-partite
matchings, $W_M^{(d)}=M^{d-1}$ and $\E[C_M^{(d)}]\sim M^{2-d}$, while for
simple $d$-matchings, $W_N^{(d)}=\binom{N-1}{d-1}$ and $\E[L_N^{(d)}]\sim
(d-1)!N^{2-d}$. We will therefore be interested in computing the (finite)
quantities
\begin{equation}\label{eq:def}
\begin{split}
\mathcal{C}^{(d)}&=\lim_{M\to\infty}M^{d-2}\E[C_M^{(d)}],\\
\mathcal{L}^{(d)}&=\lim_{N\to\infty}\frac{N^{d-2}}{(d-1)!}\E[L_N^{(d)}].
\end{split}
\end{equation}
The factor $(d-1)!$ in the second definition is meant to reflect the different
number of hyperedges to which a given node can connect, in the $d$-partite and
simple versions (this difference is absent when $d=2$). With this convention
we will find the equality $\mathcal{C}^{(d)}=d\mathcal{L}^{(d)}$, where the
remaining $d$ factor merely comes from the fact that the total number of nodes
is $N$ for simple $d$-matchings, but is $dM$ for $d$-partite matchings.

\subsection{Annealed
approximation}\label{sec:annealed}

When energies are extensive in the size $N$ of the system, the equilibrium
properties of a statistical physics model are entirely encoded in the
partition function, $Z_N(\beta)=\sum_\mathcal{M}e^{-\beta E(\mathcal{M})}$,
or, equivalently, in its logarithm, the free energy
$F_N(\beta)\equiv-\log[Z_N(\beta)]/\beta$. The free energy depends on the
realization of the elementary costs, but it is expected to be a self-averaging
quantity, {\it i.e.}, such that the free-energy density
$f(\beta)=\lim_{N\to\infty}F_N(\beta)/N$ exists and is independent of the
sample. The self-averaging property is proved for $d=2$~\cite{Aldous90}, and
we assume here that it holds for $d\geq 3$ as well. The value of the optimal
matching is given by the ground state energy, obtained as
$\lim_{\beta\to\infty}f(\beta)$, where the free energy is calculated by
performing a {\it quenched} average of the partition function, $\E[\ln Z]$,
with $\E[\cdot]$ referring to the average with respect to the realization of
the elementary costs.

A much simpler calculation is the {\it annealed} average, $\ln \E[ Z]$. Due to
the concavity of the logarithm, it yields a lower bound on the correct
quenched free energy, $f_{\rm an}(\beta)\equiv-\ln\E[Z]/(N\beta)\leq -\E[\ln
Z]/(N\beta)\equiv f(\beta)$. In fact, since the entropy
$s(\beta)=\beta^2\partial_\beta f(\beta)$ is necessarily positive for a system
with discrete degrees of freedom, the free energy $f(\beta)$ must be an
increasing function, and a tighter lower bound can be inferred for the
ground-state energy~\cite{VannimenusMezard84},
\begin{equation}
\lim_{\beta\to\infty}f(\beta)\geq \sup_{\beta>0}f_{\rm
an}(\beta).
\end{equation}

These considerations are made under the hypothesis that the energies, or
equivalently the temperature $\beta$, are correctly scaled with $N$, so that
$\lim_{\beta\to\infty}f(\beta)$ is indeed finite. Reciprocally, requiring the
annealed free energy to be extensive provides us with the appropriate scaling
of $\beta$. For $d$-index matching problems, we have
\begin{equation}
\E[Z]=\E\left[\sum_{\{n_a\}} e^{-\beta\sum_a c_a
n_a}\right]=(\#\mathcal{M})\E[e^{-\beta c_a}]^{\#\{a\in\mathcal{M}\}}
\end{equation}
where $\#\mathcal{M}$ denotes the total number of
possible matchings and $\#\{a\in\mathcal{M}\}$ the
number of hyperedges contained in a given matching.
For $d$-partite matchings, $\#\mathcal{M}=(M!)^{d-1}$
and $\#\{a\in\mathcal{M}\}=M$. To enforce the correct
scaling of the free energy, we anticipate a rescaling
in temperature of the form
$\beta=M^\alpha\hat{\beta}$, yielding
\begin{equation}\label{eq:annealed}
\ln\E[Z]=[d-1-\alpha]M\ln M -
[\ln\hat{\beta}+d-1]M+o(M).
\end{equation}
An extensive annealed free energy is therefore
obtained by taking
$\alpha=d-1$, in which case
\begin{equation}\label{eq:fannealed}
f^{(d{\rm-part)}}_{\rm
an}(\hat{\beta})=\frac{\ln\hat{\beta}
+d-1}{\hat{\beta}}.
\end{equation}
The scaling $1/(M\hat{\beta})\sim M^{2-d}$ we obtain 
corresponds to one introduced in Eq.~(\ref{eq:def}).
For simple $d$-index matchings,
$\#\mathcal{M}=N!/[(N/d)!(d!)^{N/d}]$ and
$\#\{a\in\mathcal{M}\}=N/d$. Rescaling the temperature
as
$\beta=N^{d-1}\hat{\beta}$, we get
\begin{equation}
\ln\E[Z]=-[\ln\hat{\beta}+d-1-\ln(d-1)!]N/d+o(N).
\end{equation}
To make contact with the $d$-partite case however, we adopt a slightly
different scaling, $\beta=N^{d-1}\tilde{\beta}/(d-1)!$, so that
\begin{equation}
f^{(\rm simple)}_{\rm
an}(\tilde{\beta})=\frac{\ln\tilde{\beta}
+d-1}{d\tilde{\beta}}=\frac{1}{d}f^{(d{\rm-part})}_{\rm
an}(\hat{\beta}=\tilde{\beta}).
\end{equation}
This annealed calculation illustrates the correspondence between the
$d$-partite and simple $d$-matchings stated in the previous section. Apart for
the trivial factor $d$, corresponding to the relation $N=dM$, the equality is
obtained by normalizing differently $\hat{\beta}$ and $\tilde{\beta}$, thereby
accounting for the difference in the number of hyperedges a given node locally
sees [extra factor $(d-1)!$ in Eq.~(\ref{eq:def})]. The annealed free energy
is a concave function with maximum for $\hat{\beta}_d^*=e^{2-d}$ so that we
get lower bounds $\mathcal{C}^{(d)}\geq e^{d-2}$ and $\mathcal{L}^{(d)}\geq
e^{d-2}/d$.

\subsection{Statistical physics reformulation}

From now on, we will cease distinguishing between $d$-partite and simple
$d$-matchings, and consider a unique statistical physics model that describes
both problems in a common framework. Our approach is indeed based on the
cavity method~\cite{MezardParisi01} for which only the local properties at the
level of each node are relevant, and we have seen that by making the
appropriate scalings of $\beta$, we can match the local properties of both
models. The Hamiltonian we consider is
\begin{equation}\label{eq:h}
\mathcal{H}[\{n_a\}]=\sum_a \xi_a n_a,
\end{equation}
with the $\xi_a\equiv M^{d-1} c_a$ uniformly distributed in $[0,M^{d-1}]$ for
the $d$-partite case, and $\xi_a\equiv N^{d-1} c_a/(d-1)!$ in
$[0,N^{d-1}/(d-1)!]$ for the simple case. The (inverse) temperature, denoted
by $\beta$ to simplify, will correspond to $\hat{\beta}$ for the $d$-partite
case and $\tilde{\beta}$ for the simple case. The only remaining difference
kept is the factor $d$ between the two free energies, accounting for the
relation $N=dM$. Unless explicitly stated, the formulae to be given hold for
the simple version ; to get the $d$-partite counterparts, one has consequently
to multiply by $d$ the intensive quantities.

\section{Replica symmetric solution}

The approach we adopt to treat the $d$-matching problems is the cavity method
recently developed to solve statistical physics models defined on finite
connectivity graphs \cite{MezardParisi01}. This section explains the formalism
of the replica symmetric solution for general $d$. While the correctness of
the replica symmetric approach is a mathematical fact when $d=2$, we show that
it leads to some inconsistency when $d=3$, requiring replica symmetry to be
broken.

\subsection{From complete to dilute graphs}

The hypergraph on which an instance of the simple $d$-matching problem is
defined is {\it complete}, in the sense that every possible hyperedge arises
once and is given a random cost. However, the factor nodes with the smallest
elementary costs are more likely to belong to the optimal matching; for
instance the probability that the $k$-th most costly hyperedge originating
from a given node will be included in the optimal 2-matching is
$2^{-k}$~\cite{Aldous01}. This suggests that hyperedges with large costs can
be ignored while retaining most of the structure relevant to the determination
of the optimal matching. Eliminating hyperedges results in a {\it diluted}
hypergraph, where each node is connected to only a restricted number of
hyperedges. From this point of view, in spite of being defined on a complete
graph, random matchings are effectively closer to statistical physics models
defined on finite connectivity random graphs. In fact, such a feature already
transpired from the initial replica treatment~\cite{MezardParisi85} of
$2$-index matchings where all the multioverlaps $Q_{a_1\dots a_p}$ were
required, and not only the two replica overlaps $Q_{a_1a_2}$, like in usual
Curie-Weiss mean field models of disordered systems \cite{MezardParisi87b}.

To exploit the underlying diluted structure, one possible method is to
introduce a cut-off $C$, suppress all nodes with rescaled cost $\xi_a>C$,
solve the matching problem on the diluted hypergraph, and finally send
$C\to\infty$. The hypergraph obtained by this procedure is Poissonian: if
$\xi_1,\xi_2,\dots$ are the costs ordered in increasing sequence of the
hyperedges connected to a given node, the probability for the connectivity to
be $k$ is
\begin{equation}
p_k={\rm
Prob}[\xi_1<\cdots<\xi_k<C<\xi_{k+1}<\cdots]=\binom{W_\Lambda^{(d)}}{k}\left(\frac{C}{W_\Lambda^{(d)}}\right)^k\left(1-\frac{C}{W_\Lambda^{(d)}}\right)^{W_\Lambda^{(d)}-k}\to
\frac{C^k}{k!}e^{-C},
\end{equation}
with $W_\Lambda^{(d)}$ giving the number of hyperedges to which a node is
connected, as in Sec.~\ref{sec:annealed}. Diluting the complete graph has a
major drawback however: the diluted hypergraph typically does not allow any
matching at all, since for instance there is always a finite probability
$e^{-C}$ that a given node is isolated.

To circumvent this problem, we come back to the model on the complete graph
and start by weakening the constraints, allowing a node not to belong to a
matching, at the expense of paying an extra cost. In more physical terms, we
view a matching as the close-packing limit of a lattice gas model whose
particles are subject to hard-core interactions: particles can occupy the
hyperedges but two hyperedges connected through a node can not both admit a
particle. We introduce a {\it grand-canonical} Hamiltonian
\begin{equation}\label{eq:hprime}
\mathcal{H}_\mu[\{n_a\}]=\sum_a \xi_a n_a-d\mu\sum_a
n_a=\sum_a (\xi_a-d\mu)
n_a.
\end{equation}
where $d\mu$ is a chemical potential per hyperedge
($\mu$ per node). In the limit $\mu\to\infty$, the
maximum number of hyperedges is occupied by a particle
and we recover the matching problem. For finite $\mu$
however, the constraints reflecting the hard-core
repulsion are
\begin{equation}\label{eq:soft}
\forall i,\quad \sum_{a\ :\ i\in a}n_a\leq 1,
\end{equation}
to be compared with the hard constraints of Eq.~(\ref{eq:constraint}),
recovered only in the $\mu\to\infty$ limit. Each value of $\mu$ defines an
optimization problem whose minimum energy $E_\mu$ corresponds to a zero
temperature limit $\beta\to\infty$. The solution of the matching problem thus
appears as the result of a double limit, $\beta\to\infty$ and $\mu\to\infty$.
The point is that a diluted structure is now naturally associated with the
system at finite $\mu$. Indeed since $\mathcal{H}_\mu$ is minimized by taking
$n_a=0$ whenever $\xi_a>d\mu$, the ground state is unaffected if all
hyperedges $a$ with $\xi_a>d\mu$ are suppressed, yielding the Poissonian
hypergraph considered above with $C=d\mu$. This construction allows us to
formulate the initial MIMP as the limit $\mu\to\infty$ of optimization
problems defined on Poissonian graphs with increasing mean connectivity
$d\mu$. We will give in Sec.~\ref{sec:t=0} an alternative construction based
on regular graphs.

\subsection{Cavity method}

The problem at finite $\mu$ defined on a Poissonian hypergraph can be studied
by means of the cavity method as developed for finite connectivity
graphs~\cite{MezardParisi01}; one of the main advantages of this method over
the replica method~\cite{MezardParisi85} or previous versions of the cavity
method \cite{MezardParisi86} is that it allows for a practical investigation
of replica symmetry breaking (RSB). Since we are interested in the
ground-state properties, the cavity method directly at zero temperature seems
particularly well suited~\cite{MezardParisi03}. However, it will turn out to
be necessary to get the finite temperature equations as well, and we therefore
work at finite $\beta$, postponing the discussion of the $\beta\to\infty$
limit to the next section.

\begin{figure}
\begin{minipage}[t]{.46\linewidth}
\centering
\epsfig{file=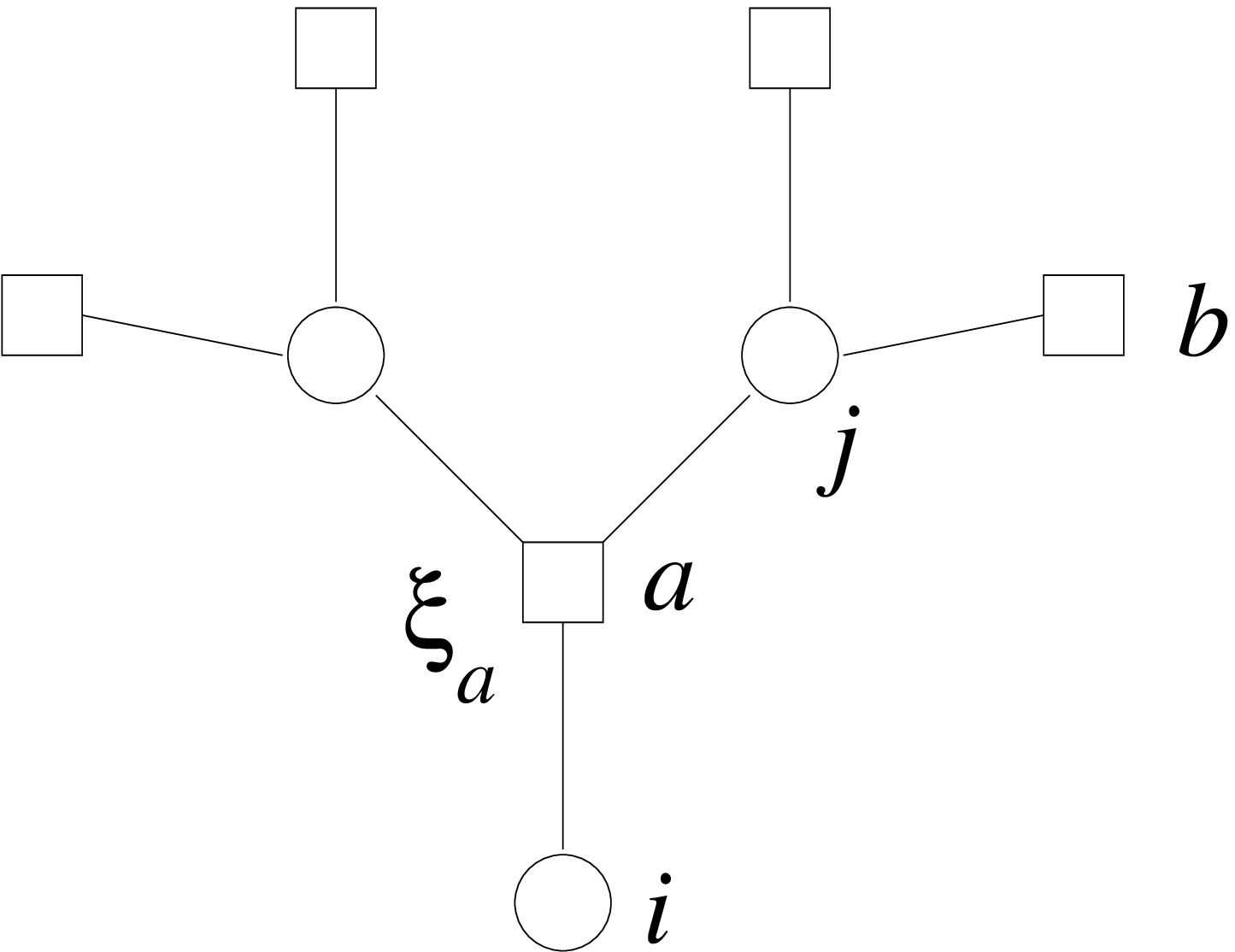,width=.7\linewidth}
\caption{\small Local structure of a Poissonian
hypergraph. When the node $i$ is removed, it leaves a
rooted-tree with root $a$.} \label{fig:trees}
\end{minipage} \hfill
\begin{minipage}[t]{.46\linewidth}
\epsfig{file=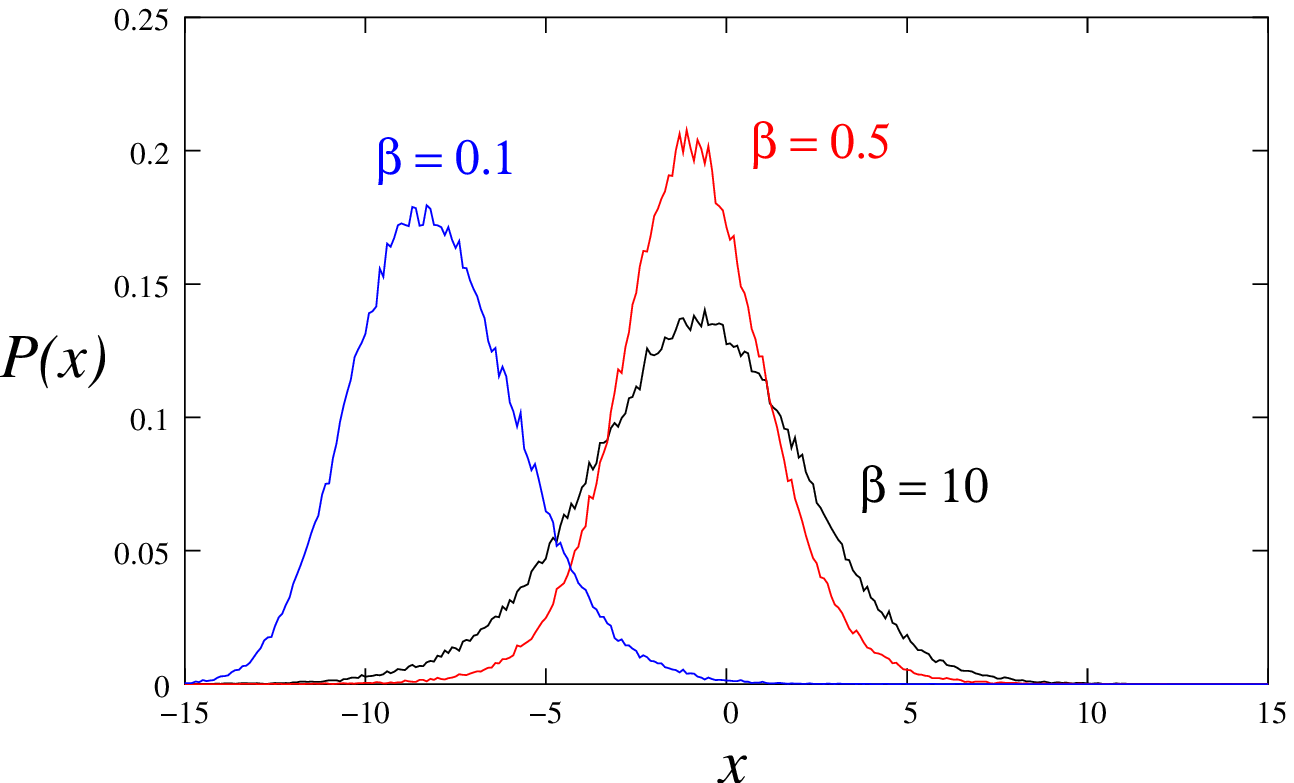,width=.9\linewidth}
\caption{\small Distribution $\mathcal{P}(x)$  of
cavity fields for 3-index matchings at different temperatures
$\beta$ in the replica symmetric approximation. These
distributions are obtained by population dynamics, using algorithm P 
described in Appendix~A, with parameters $C=60$, $\N_{\rm iter}=1$ and
$\N_{\rm pop}=200000$.} \label{fig:distr3d}\label{fig:histo3drs}
\end{minipage}
\end{figure}

In strong analogy with Aldous' framework (see Sec.~\ref{sec:t=0}), the RS
cavity method associates the diluted hypergraph with an {\it infinite tree}
or, stated differently, a tree with {\it self-consistent boundary conditions}.
The starting point is however finite rooted trees, that is trees with a
singularized node $i$ called the root. Consider for instance the part of a
tree represented in Fig.~\ref{fig:trees}: given a hyperedge $a$ and one of its
connected nodes $i$ (relation noted $i\in a$), we call $Z^{(a\to i)}$ the
partition function of the system defined on the rooted-tree with root $a$
resulting from the removal of $i$. To express it in terms of the partition
functions $Z^{(b\to j)}$ where $j$ refers to the nodes, connected to $a$, but
distinct from $i$ (noted $j\in a-i$), we decompose $Z^{(a\to i)}$ as $Z^{(a\to
i)}=Z_0^{(a\to i)}+Z_1^{(a\to i)}$, where $Z^{(a\to i)}_0$ and $Z^{(a\to
i)}_1$ are the conditional partition functions where the root $a$ is either
constrained to be empty or occupied by a particle. As an intermediate stage in
the recursion, we also introduce $Y^{(j\to a)}_0$ and $Y^{(j\to a)}_1$, which
are defined similarly to $Z^{(a\to i)}_0$ and $Z^{(a\to i)}_1$, but for
rooted-trees whose root is the node $j$, in absence of the hyperedge $a$: the
index 1 means that $j$ is already matched and the index 0 that it is not. With
the notations of Fig.~\ref{fig:trees}, we have the relation
\begin{equation}\label{eq:trees}
\begin{split}
Z^{(a\to i)}_0 &=\prod_{j\in a-i}\left( Y^{(j\to
a)}_0+Y^{(j\to a)}_1\right),\\
Z^{(a\to i)}_1 &=e^{-\beta (\xi_a-d\mu)}\prod_{j\in
a-i}Y^{(j\to a)}_0,\\
Y^{(j\to a)}_0 &=\prod_{b\in j-a}Z^{(b\to j)}_0,\\
Y^{(j\to a)}_1 &=\sum_{b\in j-a}Z^{(b\to
j)}_1\prod_{c\in j-\{a,b\}}Z^{(c\to j)}_0.
\end{split}
\end{equation}
These formulae have simple interpretations: for instance, the first line means
that when $a$ is empty, the neighboring nodes $j\in a-i$ can be equally
matched or not with upstream hyperedges, while the second line means that when
$a$ is occupied, it generates a cost $\xi_a-d\mu$ and requires the nodes $j\in
a-i$ to not be matched.
From the conditional partition functions, we define
the cavity fields
\begin{equation}
\begin{split}\label{eq:defields}
e^{\beta x^{(j\to a)}}&\equiv
e^{\beta\mu}\frac{Y_0^{(j\to a)}}
{Y_0^{(j\to a)}+Y_1^{(j\to a)}},\\
e^{\beta u^{(a\to i)}}&\equiv
e^{\beta(\xi_a-\mu)}\frac{Z_1^{(a\to i)}}{Z_0^{(a\to
i)}}.
\end{split}
\end{equation}
These definitions are made to insure a proper scaling when $\mu\to\infty$ and
recover quantities used in previous studies for $d=2$. Note however that for
finite $\beta$, it is more natural to introduce $\psi^{(j\to a)}\equiv
\exp[\beta(x^{(j\to a)}-\mu)]$ interpreted as the probability that node $j$
not matched in the absence of $a$ (or equivalently that $j$ is associated to
$a$ in a matching); this alternative notation will turn out to be particularly
convenient when discussing the freezing phenomenon, in Sec.~\ref{sec:frozen}.
On a given rooted tree, it follows from Eq.~(\ref{eq:trees}) that the fields
attached to the different oriented edges are related by the following {\it
message-passing rules},
\begin{equation}\label{eq:recrs}
\begin{split}
u^{(a\to i)}&=\sum_{j\in a-i}x^{(j\to a)},\\
x^{(j\to
a)}&=-\frac{1}{\beta}\ln\left(e^{-\beta\mu}+\sum_{b\in
j-a}e^{-\beta(\xi_a-u^{(b\to j)})}\right).
\end{split}
\end{equation}
The limit of infinite rooted trees is taken
implicitly by considering the stationary
distribution $\P(x)$ that is assumed to result from
the repeated iteration of the message passing
relations. By definition  $\P(x)$ is a distribution of
cavity fields over the different oriented edges that
satisfies the following self-consistent equation,
called the RS cavity equation,
\begin{equation}\label{eq:cavitymatching}
\P(x^{(0)})=\E_{k,\xi}
\int\prod_{a=1}^k\prod_{j_a=1}^{d-1}dx^{(j_a)}
\P(x^{(j_a)})\delta\left(x^{(0)}-\hat{x}^{(k,\xi)}[\{x^{(j_a)}\}]\right)
\end{equation}
where the function $\hat{x}^{(k,\xi)}$ is defined
according to Eq.~(\ref{eq:recrs}) as
\begin{equation}\label{eq:hatx}
\hat{x}^{(k,\xi)}[\{x^{(j_a)}\}]\equiv-\frac{1}{\beta}\ln\left(e^{-\beta\mu}+\sum_{a=1}^ke^{-\beta(\xi-\sum_{j_a=1}^{d-1}x^{(j_a)})}\right),
\end{equation}
and the expectation $\E_{k,\xi}$ expresses the average
over the disorder, which includes both an average over
the connectivity $k$ and over the rescaled costs
$\xi$,
\begin{equation}
\E_{k,\xi}[F^{(k)}(\{\xi_a\})]\equiv\sum_{k=0}^\infty
\frac{(d\mu)^ke^{-d\mu}}{k!}\prod_{a=1}^k\left(\frac{1}{d\mu}\int_0^{d\mu}
d\xi_a \right)F^{(k)}(\xi_1,\dots,\xi_k).
\end{equation}

The RS cavity equation (\ref{eq:cavitymatching}) can be solved by a population
dynamics algorithm, whose principle is presented in Appendix A; the resulting
distribution $\mathcal{P}(x)$ for $d=3$ and different $\beta$ is shown in
Fig.~\ref{fig:histo3drs}. $\mathcal{P}(x)$ contains all the information on the
equilibrium properties and, in particular, allows one to compute the
free-energy density. It can be derived from the Bethe approximation which
produces on a given hypergraph the formula
\begin{equation}
f(\beta)=\frac{1}{N}\left[\sum_i\Delta F^{(i+a\in
i)}(\beta)-\sum_a(\ell_a-1)\Delta
F^{(a)}(\beta)\right]
\end{equation}
where $\ell_a$ is the degree of hyperedge $a$, which here is $\ell_a=d$
independently of $a$. The shifts $\Delta F^{(i+a\in i)}(\beta)$ and $\Delta
F^{(a)}(\beta)$ correspond respectively to the free-energy shift induced by
the addition of a node $i$ together with its connected hyperedges $a\in i$,
and to the free-energy shift induced by the addition of hyperedge $a$. They
are given by
\begin{equation}
\begin{split}
e^{-\beta\Delta F^{(i+a\in
i)}(\beta)}&=\frac{Y_0^{(i)}+Y_1^{(i)}}{\prod_{a\in
i}\prod_{j\in
a-i}\left(Y_0^{(j)}+Y_1^{(j)}\right)}=e^{-\beta\mu}+\sum_{a\in
i}e^{-\beta(\xi_a-\sum_{j\in a-i}x^{(j\to a)})},\\
e^{-\beta\Delta
F^{(a)}(\beta)}&=\frac{Z_0^{(a)}+Z_1^{(a)}}{\prod_{j\in
a-i}\left(Y_0^{(j)}+Y_1^{(j)}\right)}=1+e^{-\beta(\xi_a-\sum_{j\in
a}x^{(j\to a)})},
\end{split}
\end{equation}
where we introduced the analogs of the partitions
functions for rooted tree, but on the complete trees:
\begin{equation}
\begin{split}
Z^{(a)}_0 &=\prod_{j\in a}\left( Y^{(j\to
a)}_0+Y^{(j\to a)}_1\right),\\
Z^{(a)}_1 &=e^{-\beta (\xi_a-d\mu)}\prod_{j\in
a}Y^{(j\to a)}_0,\\
Y^{(j)}_0 &=\prod_{b\in j}Z^{(b\to j)}_0,\\
Y^{(j)}_1 &=\sum_{b\in j}Z^{(b\to j)}_1\prod_{c\in
j-b}Z^{(c\to j)}_0.
\end{split}
\end{equation}
Physically, $Z^{(a)}_1/(Z^{(a)}_0+Z^{(a)}_1)$ gives the probability for the
hyperedge $a$ to be included in the matching. By averaging over the
realizations of the disorder, since the mean number of hyperedges per nodes is
$\mu$, we get
\begin{equation}\label{eq:fRS1}
f_{RS}(\beta) =\E[\Delta F^{(i+a\in
i)}(\beta)]-(d-1)\mu\E[\Delta F^{(a)}(\beta)]
\end{equation}
with explicitly
\begin{equation}\label{eq:fRS2}
\begin{split}
&\E[\Delta F^{(i+a\in
i)}(\beta)]=-\frac{1}{\beta}\E_{k,\xi}\int\prod_{a=1}^k\prod_{j_a=1}^{d-1}dx^{(j_a)}
\P(x^{(j_a)})\ln\left(e^{-\beta\mu}+\sum_{a=1}^ke^{-\beta(\xi_a-\sum_{j_a=1}^{d-1}x^{(j_a)})}\right),\\
&\E[\Delta F^{(a)}(\beta)]
=-\frac{1}{\beta}\E_{\xi}\int\prod_{j=1}^d
dx^{(j)}\P(x^{(j)})\ln\left(1+e^{-\beta(\xi_a-\sum_{j=1}^d
x_j)}\right).
\end{split}
\end{equation}

\subsection{Integral relations}

The $\mu\to\infty$ limit can be taken explicitly. The corresponding equations
generalize the formulae established by M\'ezard and Parisi in their first
treatment of the 2-index matching problem~\cite{MezardParisi85}. For general
$d$, they are
\begin{equation}\label{eq:Grec}
\begin{split}
G(l)&=\frac{1}{\beta}\int_{-\infty}^{+\infty}\prod_{j=1}^{d-1}dy_je^{-G(y_j)}B_d\left(l+\sum_{j=1}^{d-1}y_j\right),\\
B_d(x)&
\equiv\sum_{p=1}^\infty\frac{(-1)^{p-1}p^{d-2}e^{px}}{(p!)^d}.
\end{split}
\end{equation}
Given $G(\ell)$, the energy $\epsilon(\beta)$ and
entropy
$s(\beta)$ are
\begin{equation}\label{eq:eands}
\begin{split}
\epsilon(\beta)&=\frac{1}{\beta
d}\int_{-\infty}^{+\infty}dlG(l)e^{-G(l)},\\
s(\beta)&=\int_{-\infty}^{+\infty}dl\left[e^{-e^l}-e^{-G(l)}\right]-\frac{d-2}{d}\int_{-\infty}^{+\infty}dlG(l)e^{-G(l)},
\end{split}
\end{equation}
and the free energy is obtained as $f(\beta)=\epsilon(\beta)-s(\beta)/\beta$.
The relation between the function $G(l)$ and the order parameter
$\mathcal{P}(x)$ is, up to a change of variable, a Laplace transform,
\begin{equation}
e^{-G(l)}=\int_{-\infty}^{+\infty}
dx\mathcal{P}(x)e^{-e^{l-\beta x}}.
\end{equation}
From the practical point of view of numerically
solving the cavity 
equations, the finite $\mu$ cavity equations are
however easier to
handle than these compact formulae.

\subsection{Zero temperature limit}\label{sec:t=0}

In view of an extension of the mathematical approach from 2-index to $d$-index
matchings with $d>2$, it is interesting to discuss in some details the
relations between our equations and those used by Aldous in his rigorous study
of the 2-index matching problem~\cite{Aldous01}. Aldous' formalism is obtained
from our RS cavity equations by taking the zero temperature, $\beta\to\infty$.
When $\beta\to\infty$, Eqs.~(\ref{eq:recrs}) become
\begin{equation}
\begin{split}
u^{(a\to i)}&=\sum_{j\in a-i}x^{(j\to a)},\\
x^{(j\to a)}&=\min_{b\in j-a}(\xi_b-u^{(b\to j)}).
\end{split}
\end{equation}
Taking $\mu=\infty$ and $d=2$ leads to the
{\it recursive distributional
equation}~\cite{AldousBandyopadhyay04},
\begin{equation}\label{eq:cavity}
x^{(a)}=\min_b\left(\xi_b-x^{(b)}\right)
\end{equation}
on which Aldous' work is based~\cite{Aldous01}. A difference is however that
its costs $\xi_b$ derive from a Poisson point process (the uniform
distribution does not make sense when $\mu=\infty$). This Poisson process can
nonetheless be related to our formalism by implementing a variant of the
cut-off procedure. Consider selecting at each step of the cavity recursion the
$k$ parents of smallest costs, $k$ being now fixed. Then the successive $k$
costs are distributed according to a Poisson process with rate one.
Nonetheless, while the cavity recursion is perfectly well defined, the
corresponding system on a given hypergraph does not make sense: a hyperedge
may belong to the the list of the hyperedges with the $k$-th smallest costs
for one of its node but not for an other one. This is why we introduced the
version with a cut-off on the costs, which constitutes for finite $\mu$ a
perfectly sensible statistical physics model. From a purely formal point of
view the version with cut-off on the number of connected clauses works as
well, and provides an alternative formulation for numerically solving the
cavity equations (see Appendix A for the details and Fig.~\ref{fig:f4drs} for
an illustration).

The cavity fields at zero temperature have an interpretation in terms of
differences in ground-state energies. The cavity field $x^{(j\to a)}$
corresponds to the extra cost of a particle on node $j$ with respect to no
particle, in the absence of hyperedge $a$, and the cavity bias $u^{(a\to i)}$
to the cost of connecting the node $i$ to the hyperedge $a$. Note that these
quantities are actually well defined only if $\mu$ is kept finite, otherwise a
particle cannot be removed or added without destroying the perfect matching,
{\it i.e.}, without leaving the space of admissible configurations. Similarly,
the total fields on the complete graph are
\begin{equation}
\begin{split}
U^{(a)}&=\sum_{j\in a}x^{(j\to a)},\\
X^{(i)}&=\min_{b\in i}(\xi_b-u^{(b\to i)}).
\end{split}
\end{equation}

From the interpretation given, it appears that the hyperedges which indeed
participate to the optimal matching are those which achieve the minima, {\it
i.e.}, the solution is given by
\begin{equation}\label{eq:inclusion}
n_a =\delta_{a,a^*}, \quad a^*= \arg \min_a
(\xi_a-u^{(a\to i)}).
\end{equation}
Since this has to hold for all $i\in a$, the question arises whether this
prescription effectively defines a matching, {\it i.e.}, whether $\arg \min_a
(\xi_a-u^{(a\to i)})=\arg \min_a (\xi_a-u^{(a\to j)})$ for all $i,j\in a$. A
positive answer is obtained by generalizing to $d>2$ the {\it inclusion
criterion} invoked by Aldous when $d=2$, which states
\begin{equation}
a^*= \arg \min_a (\xi_a-u^{(a\to i)})=1\quad \iff
\quad \xi_a\leq u^{(a\to i)}+x^{(i\to a)}.
\end{equation}
The independence on $i$ is then a consequence of the
dentity $u^{(a\to i)}+x^{(i\to a)}=U^{(a)}$. The
proof of the inclusion criterion itself is
straightforward with the present notations: if $a^*=
\arg \min_a (\xi_a-u^{(a\to i)})$,
\begin{equation}
\xi_{a^*}-u^{(a^*\to i)}=\min_{b\in i}(\xi_b-u^{(b\to
i)})
\leq\min_{b\in i-a^*}(\xi_b-u^{(b\to i)})=x^{(i\to
a^*)}.
\end{equation}
Reciprocally, if $a\neq \arg \min_b (\xi_b-u^{(b\to
i)})$,
\begin{equation}
\xi_a-u^{(a\to i)}\geq\min_{b\in i}(\xi_b-u^{(b\to i)}) =\min_{b\in
i-a}(\xi_b-u^{(b\to i)})=x^{(i\to a)}.
\end{equation}

As an alternative to the Bethe formula, the value of
the optimal matching can be obtained by inferring
$\langle \xi_{a^*} \rangle$ from the distribution of
the fields $\P(x)$. Thanks to the inclusion criterion,
we have
\begin{equation}
\begin{split}
\mathcal{L}^{(d)}_{RS}=\frac{1}{d}\langle
\xi_{a^*}\rangle&=\frac{1}{d} \int_0^\infty d\xi\ \xi\
{\rm Prob}(U>\xi)\\
&=\frac{1}{d} \int_0^\infty d\xi\int\prod_{j=1}^d
dx_j\P(x_j)\ \xi\
\theta\left(\sum_{j=1}^dx_j-\xi\right)
\end{split}
\end{equation}
where the factor $d$ corresponds to the number of nodes
per hyperedge and $\theta$ represents the Heaviside
function, $\theta(x)=1$ if $x>0$ and 0 otherwise.
The RS cavity equations at zero temperature can also
be written in terms of closed integral relations that
generalize known equalities for the $d=2$ case,
\begin{equation}
\tilde{G}(x)=\int_{\sum_jt_j>-x} \prod_{j=1}^{d-1}
dt_j\tilde{G}'(t_j)e^{-\tilde{G}(t_j)}
\left(x+\sum_{j=1}^{d-1}
t_j\right),
\end{equation}
\begin{equation}\label{eq:energyinclusion}
\mathcal{L}^{(d)}_{RS} =
\frac{1}{2d}\int_{\sum_jx_j>0}
\prod_{j=1}^ddx_j\tilde{G}'(x_j)e^{-\tilde{G}(x_j)}\left(\sum_{j=1}^dx_j\right)^2.
\end{equation}
The distribution $\tilde{G}(x)$ is related to the RS
distribution $\mathcal{P}$ of the cavity fields by
\begin{equation}
\tilde{G}(x)=-\ln\int_x^\infty dt \mathcal{P}(t),
\end{equation}
and can be obtained from the finite temperature order parameter
$G(l)=G_\beta(l)$ given in Eq.~(\ref{eq:Grec}) by
\begin{equation}
\tilde{G}(x)=\lim_{\beta\to\infty}G_\beta(\beta^{1/(d-1)}x).
\end{equation}
Comparing with the predictions of the cavity method,
$\mathcal{L}^{(d)}_{RS} = \E[\Delta\epsilon^{(i+a\in
i)}]-(d-1)\mu\E[\Delta\epsilon^{(a)}]$, we obtain the
consistency condition that the RS distribution
must satisfy
\begin{equation}
\E[x]=\frac{2-d}{2d}\E\left[\left(\sum_{j=1}^dx_j\right)^2\theta\left(\sum_{j=1}^dx_j\right)\right]
\end{equation}
where the average $\E[\cdot]$ is here taken with respect to $\mathcal{P}$.
This formula is indeed numerically verified with a good precision, in
agreement with the equivalence between the two approaches. As a corollary, it
shows that $\E[x]<0$ unless $d=2$ where $\E[x]=0$ (we recall that in this case
one has in fact an explicit formula \cite{MezardParisi85},
$\mathcal{P}(x)=1/[4\cosh^2(x/2)]$). Finally, we note that for $d>2$, the RS
energy at zero temperature is only dependent on the mean of $\mathcal{P}(x)$,
$\mathcal{L}^{(d)}_{RS}=-\E[x]/(d-2)$. However as shown in the following, the
RS approach yields incorrect predictions when $d\geq 3$.

\begin{figure}
\begin{minipage}[t]{.46\linewidth}
\centering
\epsfig{file=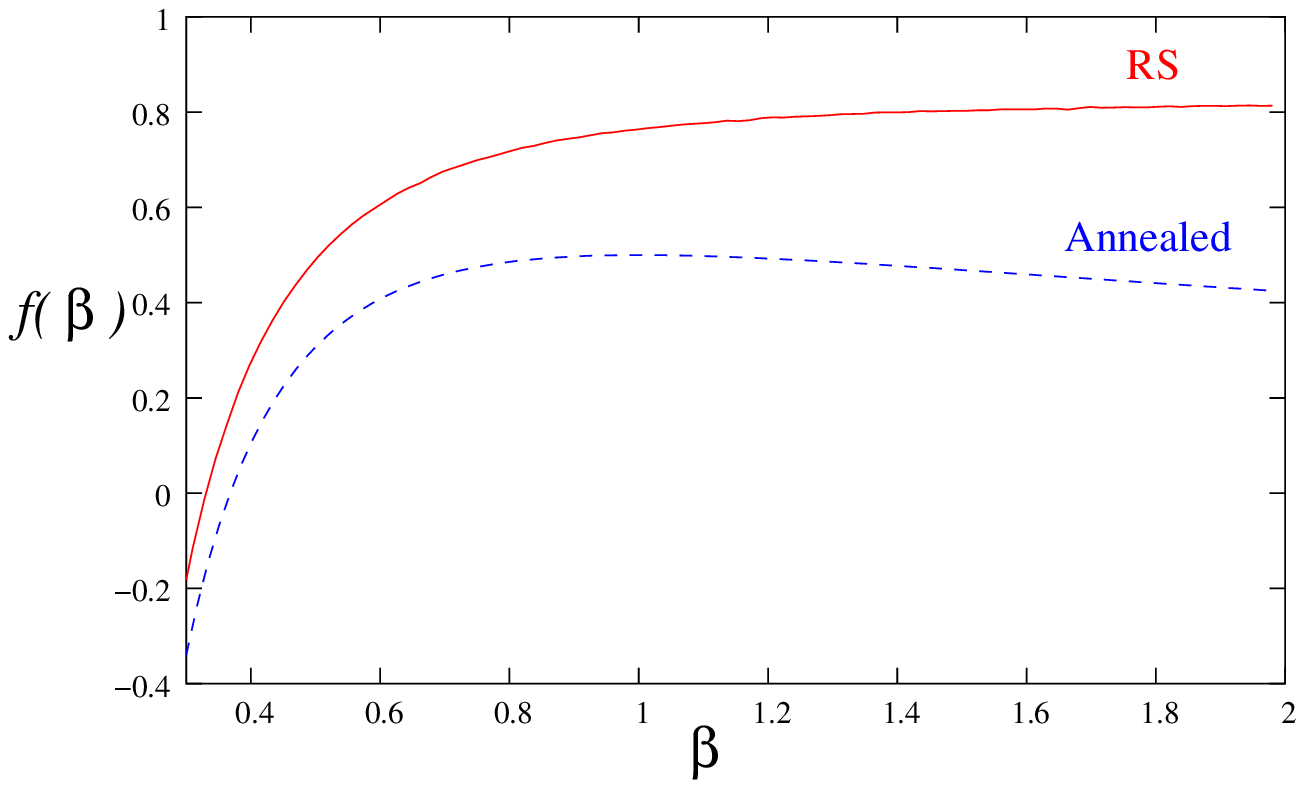,width=\textwidth}
\caption{\small Annealed and replica symmetric free
energies for the simple 2-index matching problem.
While the annealed curve has a maximum at 1/2, the RS
curve is strictly monotone. Its $\beta\to\infty$ limit
corresponds to the exactly known value $\pi^2/12\simeq 0.82$.}
\label{fig:frs2d}
\end{minipage} \hfill
\begin{minipage}[t]{.46\linewidth}
\centering
\epsfig{file=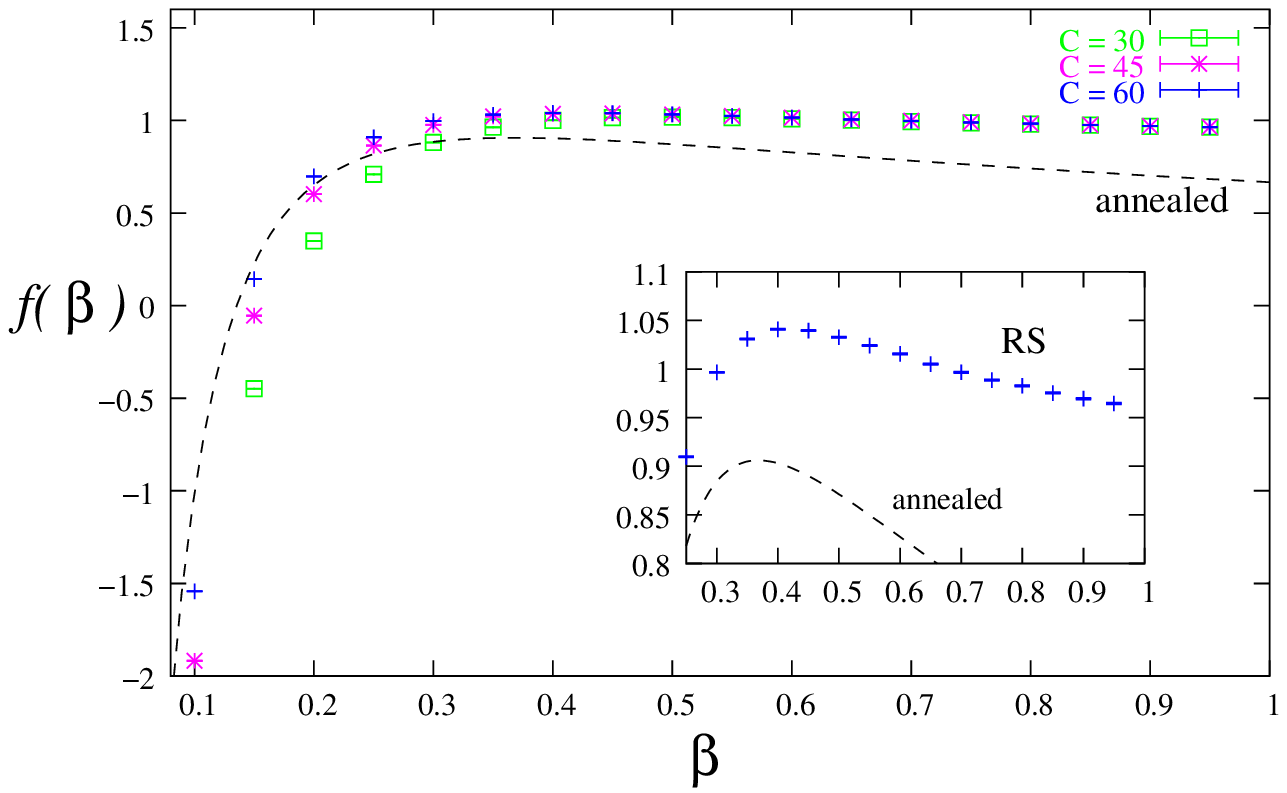,width=\textwidth}
\caption{\small Replica symmetric free energy
$f_{RS}(\beta)$ for the 3-index matching problem with
different cut-offs $C=d\mu=30,45,60$ as a function of
the inverse temperature $\beta$. This curve has been obtained using algorithm P
of Appendix~A with parameters $\N_{\rm pop}= 20 000$
and $\N_{\rm iter}= 20 000$; for
comparison, the annealed free energy is also
represented with a dashed line. In inset is a zoom of
the data with $C=60$ more clearly displaying the
non-physical decrease of $f_{RS}(\beta)$ for $\beta>\beta_s\simeq
0.41$.}
\label{fig:frs3d}
\end{minipage}
\end{figure}

\subsection{Entropy crisis}

\begin{figure}
\begin{minipage}[t]{.46\linewidth}
\centering
\epsfig{file=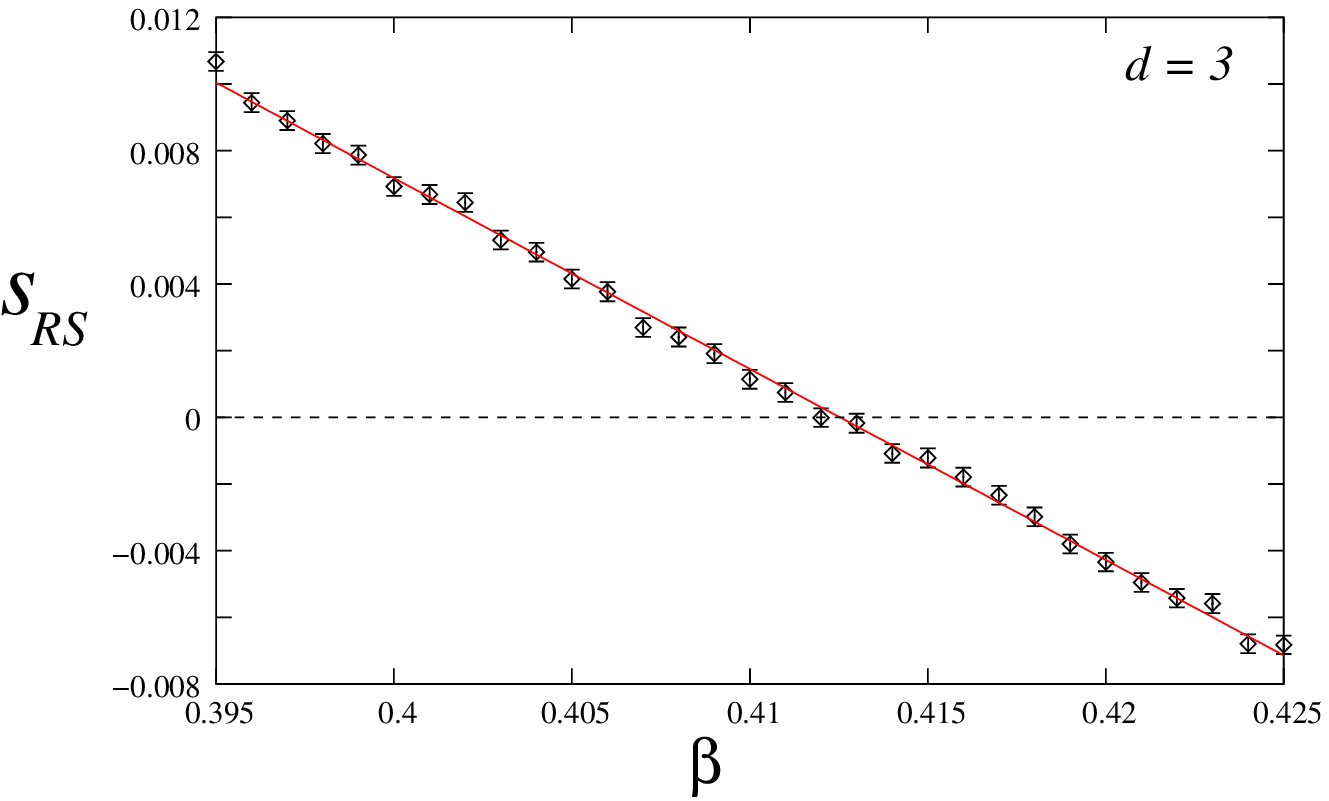,width=\linewidth}
\caption{RS entropy of the 3-index matching problem. The data is from
population dynamics, using algorithm R presented in Appendix~A, with $K=50$,
$\N_{\rm pop}= 50000$ and $\N_{\rm iter}= 5000$. The line is a linear
regression. The RS entropy is found to vanish at $\beta_c=0.412\pm
0.001$}\label{fig:betac}
\end{minipage}
 \hfill
\begin{minipage}[t]{.46\linewidth}
\centering
\epsfig{file=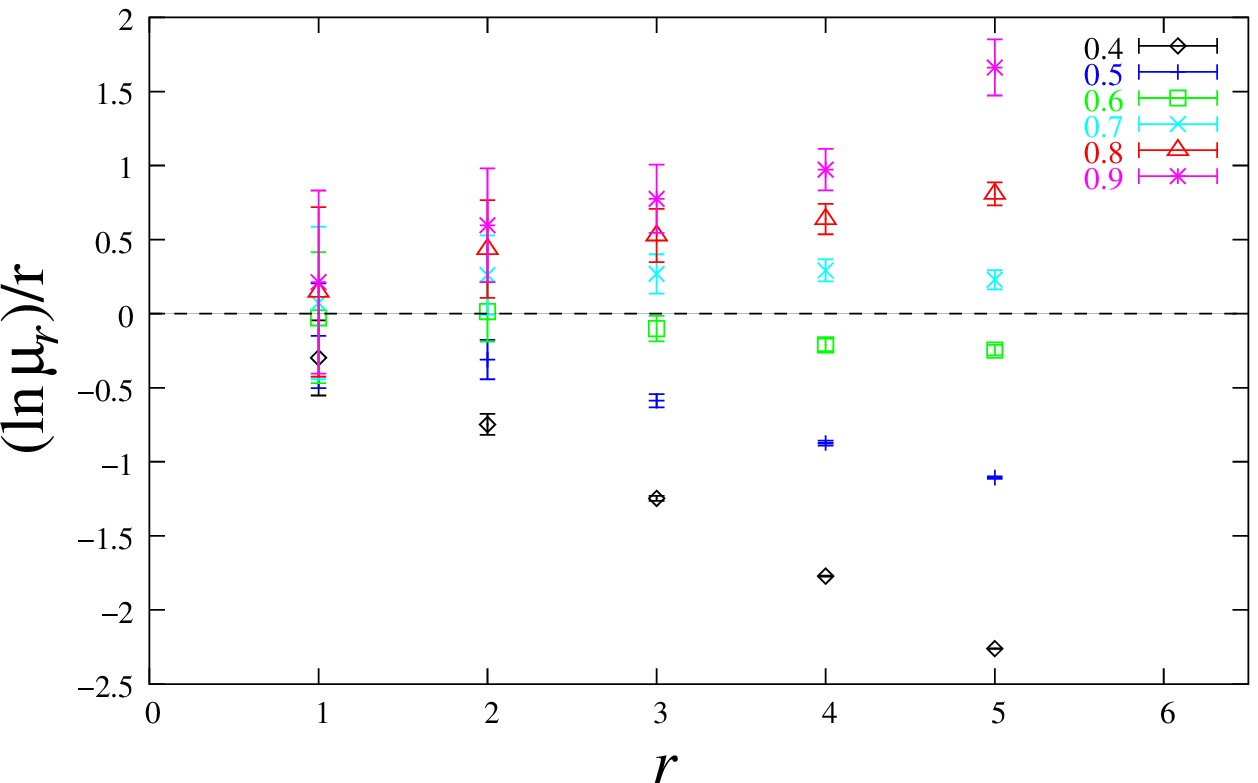,width=\linewidth}
\caption{\small Stability analysis of the RS solution
for the
3-index matching problem at finite temperature.
$(\ln\mu_r)/r$ is
plotted versus $r$ for different temperatures $\beta$
(from algorithm P given in Appendix~A with $C=36$,
$\N_{\rm pop}= 20000$ and $\N_{\rm iter}= 10^9$). The RS solution is stable
if the slope
of $(\ln\mu_r)/r$ is negative (see text), which is
found to be the
case for $\beta<\beta_i\simeq 0.6$.}
\label{fig:stab3d}
\end{minipage}
\end{figure}

Using the population dynamics algorithm described in Appendix A, we obtain for
the RS free energy $f_{RS}(\beta)$ the curves displayed in
Fig.~\ref{fig:frs2d} ($d=2$) and \ref{fig:frs3d} ($d=3$). For $d=2$, the free
energy is an increasing function of $\beta$ with limit
$f_{RS}(\beta=\infty)=\pi^2/12\simeq 0.82$ corresponding to the cost of a
minimal 2-index matching. The free energy obtained for $d=3$ is qualitatively
different, as it displays a maximum at a finite temperature $\beta_s\simeq
0.41$ (see Fig.\ref{fig:betac}). This {\it entropy crisis} reflects an
inconsistency of the RS approach~\cite{fnViolationAnnealed}. If one assumes
the RS approximation holds at high temperature in some range of temperature (a
non-trivial statement), a phase transition must occur at some
$\beta_c\leq\beta_s$.

\subsection{Stability of the replica symmetric Ansatz}

Replica symmetry fails to correctly describe the low temperature properties of
many frustrated systems~\cite{MezardParisi87b}. A necessary requirement for
its validity is that it be stable. Here we show that when $d=3$ the RS
solution is unstable below a strictly posisive temperature, that is for
$\beta>\beta_i$. Even if the breakdown of the RS hypothesis was already
inferred above from the negative value of the RS entropy, studying the
stability is instructive since the relative positions of $\beta_i$ and
$\beta_s$ will establish the discontinuous nature of the phase transition.
In~\cite{MezardParisi87}, M\'ezard and Parisi used the replica method to prove
that the RS Ansatz is stable when $d=2$~\cite{MezardParisi85}; their approach
is however quite complicated (see~\cite{ParisiRatieville01} for a recent
reexamination of their analysis), and to tackle the $d=3$ case, we adopt a
simpler approach based on the cavity method~\cite{MontanariRicci03}.
Physically, it amounts to computing the non-linear susceptibility $\chi_2$ and
checking that it does not diverge~\cite{RivoireBiroli04}. Picking a hyperedge
labeled 0 at random, this susceptibility is written
\begin{equation}\label{eq:suscept}
\chi_2=\sum_a\langle n_0 n_a\rangle_c^2 \simeq
\sum_{r=0}^\infty [C(d-1)]^r\E[\langle n_0
n_r\rangle_c^2]
\end{equation}
where $\E[\cdot]$ denotes the thermal average and $\E[\cdot]$ the spatial
average over the disorder. Using the fluctuation-dissipation relation, the
averaged squared correlation function $\E[\langle n_0 n_r\rangle_c^2]$ between
two hyperedges separated by distance $r$ can be expressed in terms of the
cavity fields as~\cite{RivoireBiroli04}
\begin{equation}
\E[\langle n_0 n_r\rangle_c^2]\sim\E\left[
\prod_{i=1}^r\left(\frac{\partial
\hat{x}^{(k,\xi)}(x_{i_1},\dots,x_{i_{(d-1)k}})}{\partial
x_{i_1}}\right)^2\right]\quad\quad(r\to\infty),
\end{equation}
where the average $\E[\cdot]$ is performed with respect to the distribution of
the disorder $(k,\xi)$ and to the distribution $\P(x)$ of the cavity fields,
except for the $x_{i_1}$ with $i>1$ which are fixed by
$x_{(i+1)_1}=\hat{x}^{(k,\xi)}(x_{i_1},\dots,x_{i_{(d-1)k}})$. To determine
whether the series in Eq.~(\ref{eq:suscept}) converges or not, we compute
\begin{equation}
\ln \mu_r=r\ln
[C(d-1)]+\ln\E\left[ \prod_{i=1}^r\left(\frac{\partial
\hat{x}^{(k,\xi)}(x_{i_1},\dots,x_{i_{(d-1)k}})}{\partial
x_{i_1}}\right)^2\right]
\end{equation}
by using cavity fields from the population dynamics, and check whether
$\lim_{r\to\infty}(\ln\mu_r)/r<0$ or not. The numerical results are limited to
small values of $r$, but as shown in Fig.~\ref{fig:stab3d} they are sufficient
to conclude unambiguously that an instability shows up for 3-index matchings
at $\beta_i\simeq 0.6$, thus confirming the incorrectness of the RS Ansatz for
describing the $\beta=\infty$ limit (the same procedure with $d=2$
consistently finds no instability). In addition, since the instability takes
place only after the entropy crisis, $\beta_i>\beta_s$, we conclude from this
analysis that the phase transition, located at $\beta_c\leq\beta_s$, must be
{\it discontinuous} as a function of the order parameter.

\section{Replica symmetry breaking}

The inconsistencies of the RS Ansatz indicate that replica
symmetry must be broken in the low temperature phase.
This feature is present in many other NP-hard combinatorial
optimization problems and is commonly overcome by
adopting a one-step replica symmetry breaking (1RSB), which, in
most favorable cases, turns out to be exact.

\subsection{General 1RSB Ansatz}\label{sec:1RSB}

As formulated by Aldous with the {\it essential uniqueness
property}~\cite{Aldous01}, replica symmetry in matching problems means that
{\it quasi-solutions}, that is low energy configurations (LECs), all share
most of their hyperedges. In contrast, replica symmetry breaking (RSB) refers
to a situation where LECs arise, which, while being close in cost to the
optimal solution, are far apart in the configurational space (the measure of
distances is the overlap between two matchings, {\it i.e.}, the fraction of
common hyperedges, see Sec.~\ref{sec:dist}). One-step replica symmetry
breaking (1RSB) is a particular scheme of RSB where the structure of the set
of LECs can be described with only two characteristic distances, $d_0$ and
$d_1<d_0$. For it to be correct, two LECs taken at random (with the Gibbs
probability measure when working at finite $\beta$) must be typically found
either at distance $d_0$ or $d_1$. In the replica jargon, close by LECs (at
the short distance $d_1$) are said to belong to the same {\it state} (or {\it
cluster}). At the level of 1RSB, it is assumed that the number
$\mathcal{N}_N(f)$ of states with a given free energy $f$ grows exponentially
with $N$ and is characterized by a {\it complexity} $\Sigma(f)$ defined by
$\Sigma(f)=\lim_{N\to\infty} [\ln\mathcal{N}_N(f)]/N$.

The 1RSB cavity method derives this ``entropy of states'' by a Legendre
transformation method mimicking the derivation of entropy from the free energy
in canonical statistical mechanics \cite{Monasson95}. The object generalizing
the free energy is the replica potential $\phi(\beta,m)$; the parameter $m$ is
the Lagrange multiplier fixing the free energy of the relevant states, in the
same way that the temperature $\beta$ selects the energy of equilibrium
configurations in the canonical ensemble. The replica potential is defined as
\begin{equation}\label{eq:phi}
e^{-N\beta m\phi(\beta,m)}\equiv\sum_\alpha e^{-N\beta
mf_\alpha},
\end{equation}
where the sum is over the states $\alpha$, and $f_\alpha$ denotes the free
energy of a system whose configurations are restricted to $\alpha$. To obtain
the relevant states for the equilibrium properties, replica theory prescribes
to choose the $m$ in $[0,1]$ that maximizes
$\phi(\beta,m)$~\cite{MezardParisi87b}, so that the equilibrium free energy is
given by
\begin{equation}
f_{\rm 1RSB}(\beta)=\max_{0\leq m \leq
1}\phi(\beta,m).
\end{equation}
Calculating $\phi(\beta,m)$ requires introducing as order parameter a
distribution $\mathcal{Q}[Q^{(j\to a)}]$ over the oriented edges $(j\to a)$ of
distributions $Q^{(j\to a)}(x)$ of the cavity fields, taken over the different
states $\alpha$~\cite{MezardParisi01}. The 1RSB cavity equations for the order
parameter read
\begin{equation}\label{eq:1rsb}
\begin{split}
\mathcal{Q}[Q^{(0)}]&=\mathbb{E}_{k,\xi}\int\prod_{a=1}^k\prod_{j_a=1}^{d-1}
\mathcal{D}Q^{(j_a)}\mathcal{Q}[Q^{(j_a)}]\delta\left[Q^{(0)}-\hat{Q}^{(k,\xi)}[\{Q^{(j_a)}\}]\right],\\
\hat{Q}^{(k,\xi)}[\{Q^{(j_a)}\}](x^{(0)})&=\frac{1}{Z}\int
\prod_{a=1}^k\prod_{j_a=1}^{d-1}dx^{(j_a)}Q^{(j_a)}(x^{(j_a)})\delta\left(x^{(0)}-\hat{x}^{(k,\xi)}(\{x^{(j_a)}\})\right)e^{-\beta
m \Delta \hat{F}^{(k,\xi)}_n(\{x^{(j_a)}\})},
\end{split}
\end{equation}
where $\hat{x}^{(k,\xi)}$ is given by
Eq.~(\ref{eq:hatx}) and the reweighting term is
\begin{equation}\label{eq:reweight}
e^{-\beta\Delta
\hat{F}^{(k,\xi)}_n(\{x^{(j_a)}\})}=e^{-\beta\mu}+\sum_{a=1}^ke^{-\beta(\xi_a-\sum_{j_a=1}^{d-1}x^{(j_a)})}.
\end{equation}
The latter corresponds to the shift of free energy due to the addition of the
new node. Its presence insures that the different states described by the
$Q^{(j\to a)}(x)$ have indeed all the same free energy, in spite of the fact
that the addition of a node inevitably introduces a free-energy shift. The
distribution $\mathcal{Q}[Q]$ determines the replica potential $\phi(\beta,m)$
whose explicit expression is
\begin{equation}
\phi(\beta,m)=\E[\Phi^{(i+a\in
i)}(\beta,m)]-(d-1)\mu\E[\Phi^{(a)}(\beta,m)]
\end{equation}
with
\begin{equation}
\E[\Phi^{(i+a\in
i)}(\beta,m)]=-\frac{1}{\beta}\mathbb{E}_{(k,\xi)}\int\prod_{a=1}^k\prod_{j_a=1}^{d-1}\mathcal{D}Q^{(j_a)}\mathcal{Q}[Q^{(j_a)}]\ln\left[\int
\prod_{a=1}^k\prod_{j_a=1}^{d-1}dx^{(j_a)}Q^{(j_a)}(x^{(j_a)})e^{-m\beta\Delta
\hat{F}^{(k,\xi)}_n(\{x^{(j_a)}\})}\right]
\end{equation}
and
\begin{equation}
\E[\Phi^{(a)}(\beta,m)]=-\frac{1}{\beta}\mathbb{E}_\xi\int\prod_{j=1}^d\mathcal{D}Q^{(j)}\mathcal{Q}[Q^{(j)}]\ln\left[\int
\prod_{j=1}^d
dx^{(j)}Q^{(j)}(x^{(j)})\left(1+e^{-\beta(\xi-\sum_{j=1}^dx^{(j)})}\right)^m\right].
\end{equation}

The 1RSB equations can in principle be numerically solved via a
population dynamics algorithm~\cite{MezardParisi01}. However, our
efforts in this direction failed to yield a sensible
order parameter because the fields were found to diverge as $\mu$
was increased : the reason for this behavior is elucidated below.

\subsection{Frozen 1RSB Ansatz}\label{sec:frozen}

Although rarely explicitly mentioned, there exists a replica symmetry breaking
Ansatz somewhat intermediate between the RS and general 1RSB as just
described. The {\it frozen 1RSB Ansatz}, which will be argued to apply to
matchings, is a particular realization of the 1RSB scheme where states are
made of single configurations (or, more generally, of a non-exponential number
of configurations). In such a case, all the information can be extracted from
the RS quantities, provided they are adequately reinterpreted. Consider for
instance the definition given by Eq.~(\ref{eq:phi}) in the special case where
states $\alpha$ have no internal entropy, {\it i.e.}, $f_\alpha=\e_\alpha$. We
thus have
\begin{equation}
e^{-N\beta m\phi(\beta,m)}\equiv\sum_\alpha e^{-N\beta
mf_\alpha}=\sum_\alpha e^{-N\beta m\e_\alpha}\equiv
e^{-N \beta mf_{RS}(\beta m)}
\end{equation}
where the last equality holds because of the very definition of a RS free
energy. The replica potential $\phi$ can therefore be expressed in term of the
RS free energy only,
\begin{equation}
\phi(\beta,m)=f_{RS}(\beta m).
\end{equation}

Following the prescriptions of replica theory, the quenched free energy is
obtained by maximizing $\phi(\beta,m)$ over $m\in [0,1]$. Being a concave
function function, the RS free energy can have at most one maximum. If
$\beta_s$ denotes the location of this maximum (with maybe $\beta_s=\infty$,
like for 2-index matchings), we obtain that
$f_{1RSB}(\beta)=\phi(1,\beta)=f_{RS}(\beta)$ for $\beta<\beta_s$ and
$f_{1RSB}(\beta)=\phi(\beta_s/\beta,\beta)=f_{RS}(\beta_s)$ for
$\beta>\beta_s$. In other words, starting from the assumption that the content
of states is trivial, the frozen Ansatz predicts a complete freezing of the
system at the point $\beta_s$ where the RS entropy becomes zero : for
$\beta>\beta_s$, the system is trapped in a single configuration and its free
energy stays constant when the temperature is further decreased ($\beta$
increased).

This scenario is already known to apply to a few models of disordered systems,
including the random energy model (REM)~\cite{Derrida80}, the directed polymer
on disordered trees~\cite{DerridaSpohn88}, the binary
perceptron~\cite{KrauthMezard89} and the {\sc XOR}-SAT problem on its
core~\cite{MezardRicci03} (with a particular case being error-correcting codes
of the Gallager type~\cite{Montanari01}). Our intention is here both to add
the matchings to this list, and to clarify the conditions under which such a
scenario may apply. At this stage, we can already state the following
necessary conditions (all satisfied by $d$-index matchings with $d\geq 3$):

$(i)$ the RS entropy must become negative at a finite
$\beta_s$;

$(ii)$ the RS solution must be stable up to (at least)
$\beta_s$;

$(iii)$ no discontinuous 1RSB transition must be
detected before
$\beta_s$.\\
In addition to these properties, the consistency of
the frozen Ansatz requires the model to have
particular kinds of constraints, called {\it hard
constraints}. Elucidating this point requires a more
refined description of the relation between the frozen
1RSB order parameter and the RS order parameter. First
remember that in the RS picture at finite temperature
$\beta$, one has a spatial distribution $\P(x^{(j\to
a)})$ of cavity fields, where following
Eq.~(\ref{eq:defields}), $\psi_{RS}^{(j\to
a)}\equiv\exp[\beta (x^{(j\to a)}-\mu)]$ is
interpreted as giving the probability under the
Boltzmann measure that node $j$ is not matched given
that the hyperedge $a$ is absent. For a general 1RSB
problem, the order parameter is instead $\Q[Q^{(j\to
a)}(x^{(j\to a)})]$ where $\psi^{(j\to
a)}\equiv\exp[\beta (x^{(j\to a)}-\mu)]$ is again a
thermal probability, but now restricted to a
particular state taken from the distribution over
states $Q^{(j\to a)}$. In this context, a RS system,
characterized by a single state, has $Q^{(j\to
a)}(\psi^{(j\to a)})=\delta(\psi^{(j\to
a)}-\psi^{(j\to a)}_{RS})$. For a system in a frozen
glassy phase instead, the thermal averages inside each
state are trivial since there is a single frozen
configuration, $\psi^{(j\to a)}=0$ or 1 meaning that a
particle is present or absent with probability one.
Therefore, the relation with the RS order parameter
has the form
\begin{equation}
Q^{(j\to a)}(\psi^{(j\to a)})=\psi^{(j\to
a)}_{RS}\delta(\psi^{(j\to a)})+(1-\psi^{(j\to
a)}_{RS})\delta(\psi^{(j\to a)}-1).
\end{equation}
Plugging this expression into the general 1RSB cavity
equation, it is found that such an Ansatz is
consistent only if the system satisfies the condition
that in the cavity recursion, the variable on a node
is completely determined by the values of the
variables on the neighboring nodes. Such is the case
with matchings when $\mu=\infty$ where a particle is
to be assigned to a hyperedge if and only if none of
the neighboring edges are occupied. This is however
not the case in all constraint problems. Consider for
instance the 3-coloring problem where each node is
assigned one of three colors with the constraint that
its color must differ from its neighbors : in the case
where all the neighbors have the same color, the
choice is left for the node between the two other
colors. When a variable is fixed by the value of its
neighbors in the cavity recursion, we say that the
system has {\it hard constraints} ; hard constraints
can be shown~\cite{these} to indeed be present in the
binary perceptron and in the {\sc XOR}-SAT model on its
core, models where the frozen Ansatz applies
too. Finally, we note that in the presence of hard
constraints, the cavity fields  $\psi^{(j\to a)}$ take
at the 1RSB level values 0 and 1 only, which are
associated with $x^{(j\to a)}=\mu$ and $-\infty$. This
explains the divergences observed when trying to
implement the 1RSB population algorithm at zero
temperature with $\mu\to\infty$.

\subsection{Distances}\label{sec:dist}

\begin{figure}
\begin{minipage}[t]{.46\linewidth}
\centering
\epsfig{file=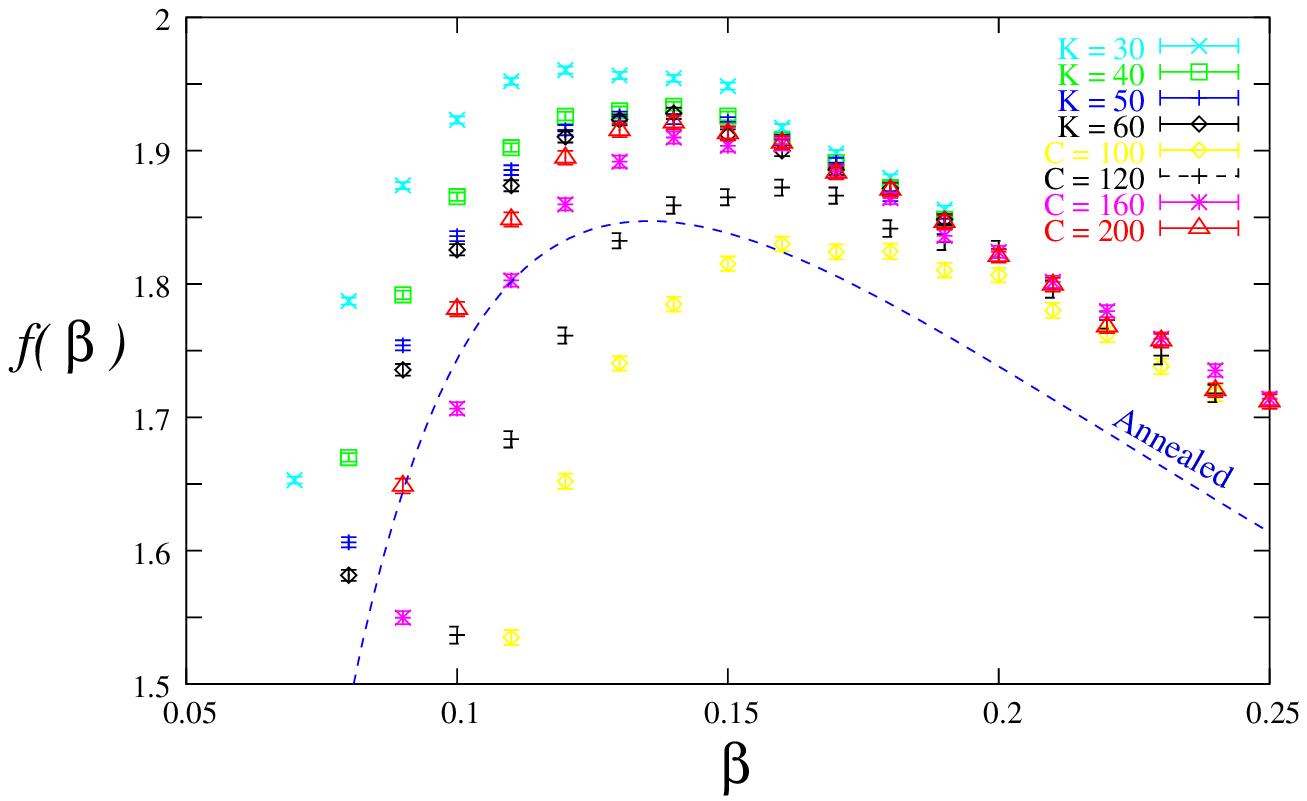,width=\linewidth}
\caption{RS free energies for $d=4$ as obtained from the 
two versions of the population dynamics algorithm described in 
Appendix~A (here with $\N_{\rm pop}=10000$ and $\N_{\rm iter}=1000$). 
Note that the approximation based on Poissonian graphs 
(mean connectivities $C=100,120,160,200$) approaches the 
solution from below, while the approximation based on regular 
graphs (fixed connectivities $K=30,40,50,60$) approaches it 
from above. As expected, the two limits $C\to\infty$ and 
$K\to\infty$ are found to match.}\label{fig:f4drs}
\end{minipage}
 \hfill
\begin{minipage}[t]{.46\linewidth}
\centering
\epsfig{file=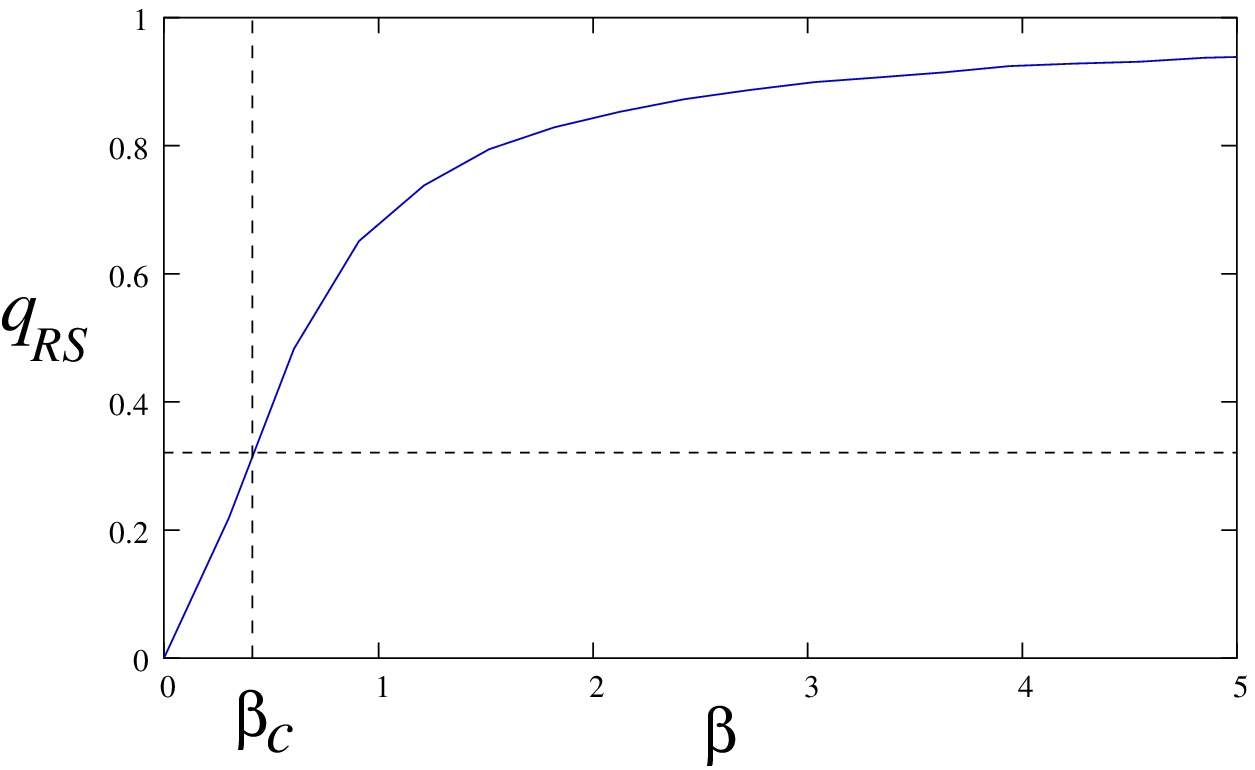,width=\linewidth}
\caption{Overlap $q(\beta)=\E[\langle n_a\rangle^2]$
in the 3-index matching as given by Eq.~(\ref{eq:q}). In
particular $q(\beta_c)=0.321\pm 0.002$ describes the
typical overlap between two low-energy matchings, that
is the fraction of hyperedges generically
share.}\label{fig:overlap}
\end{minipage}
\end{figure}

As mentioned in Sec.~\ref{sec:1RSB}, a 1RSB glassy
system is generally described by two distances, $d_0$
corresponding to the typical distance between two states, and $d_1$
corresponding to the typical distance between two configurations inside
a common state. In the case of a frozen 1RSB glassy phase, one
has however $d_1=0$ and the structure of low-energy configurations
(LECs) is characterized by only one distance, $d_0$. If $\langle
n_a\rangle$ denotes the mean occupancy of a particular hyperedge
$a$, with the average $\langle \cdot\rangle$ taken over the LECs,
the probability for $a$ to belong to two different LECs is
given by $\langle n_a\rangle^2$. Averaging over the different
hyperedges, it defines the overlap
\begin{equation}
q=\E[\langle n_a\rangle^2],
\end{equation}
which is directly related to the typical distance
between LECs through $d_0=1-q$. As argued before, for
a system in a frozen glassy phase the distribution of
energies of the LECs is described by the thermal
average at $\beta_c$ in the RS approximation, so that
\begin{equation}
\langle
n_a\rangle=\frac{Y_1^{(a)}}{Y_0^{(a)}+Y_1^{(a)}}=\frac{1}{1+e^{-\beta_c(\xi_a-\sum_{i\in
a}x^{(i\to a)})}}.
\end{equation}
Averaging over the disorder therefore yields
\begin{equation}\label{eq:q}
q=\E[\langle
n_a\rangle_{\beta_c}^2]=\E_{\xi_a}\int\prod_{j=1}^ddx^{(j)}\P(x^{(j)})\left(1+e^{-\beta_c(\xi_a-\sum_{i\in
a}x^{(j)})}\right)^{-2}.
\end{equation}
The overlap $q(\beta)$ is represented for all values
of $\beta$ in Fig.~\ref{fig:overlap} when $d=3$; given the value of
$\beta_c$ obtained before, we get $q=q(\beta_c)=0.321\pm 0.002$.

\section{Numerical analysis of finite size systems}
\label{sec:finite}

The theoretical analysis provided concerned the
$M \to \infty$ limit. How
is that limit reached, and in particular is the convergence
exponentially fast in $M$ or is it algebraic?
To answer such questions, we consider in this section
the properties of $d$-partite matchings when $M$ is
finite; in the absence of other tools, we do this
numerically. It should be clear that the most
challenging questions concern the low temperature phase 
of our system; because of that, we will focus on the 
optimum matching and low lying excitations. Even though such 
a numerical approach requires sampling the disorder (random instances)
and extracting for instance distributions with inevitable
statistical uncertainties, it will give evidence that our
frozen 1RSB Ansatz is correct; it will also provide
some statistical properties of finite size systems that
are of interest on their own.

\subsection{The branch and bound procedure}
\label{subsec:bb}
When $M$ is very small, it is possible to enumerate
all $\left[ M ! \right]^{d-1}$ $d$-partite matchings of a given sample.
Not surprisingly, this becomes unwieldy even when
$M$ reaches 10, forcing us to choose an alternate
approach. Since
it is the \emph{low energy} matchings that are of
greatest interest,
we have developed a branch and bound algorithm that
computes
the $p$ lowest energy matchings, for any given $p$.
Some technical aspects of the algorithm are presented in Appendix~B, but
the essential elements are as follows.

We represent a matching via a list of $M$
hyperedges, one for
each of the $M$ sites of the first set (recall that
there are
$d$ sets, each of $M$ sites). Such a representation
includes also some non-legal matchings as some of the
sites
in the second or higher sets could belong to more than
one hyperedge; if a matching is not legal, it is
discarded.
This representation can be mapped onto a rooted tree:
each level of the tree is associated with one of the sites
of the first set,
while a segment (branch) emerging from a node
corresponds to a
choice of hyperedge that contains the site of that
node's level. The root
node is associated with the first site, the nodes of
the next
level are associated with the second site, etc... This
tree is regular,
each node having $M^{d-1}$ outgoing segments as there are that
many hyper-edges containing a given site of the first set.
Furthermore, it has
$M+1$ levels: there is one level for each site of the
first set
while the last level consists of leaves rather than of
nodes; each
leaf corresponds
to a candidate matching specified by the list of
hyperedges
obtained when going from the tree's root to that leaf. This
list
may correspond to a legal matching or not, but each
matching
appears exactly once as a leaf. (In fact, there are
$M^{M(d-1)}$ leaves while there are only
$\left[ M ! \right]^{d-1}$ legal matchings.)

The principle of the branch and bound algorithm is to
find those leaves which satisfy the desired criterion
(the energy must be less or equal to that of the $p$th 
lowest energy matching)
by exploiting a pruning procedure, thereby avoiding
having to explore all leaves. To begin our pruned
search, we produce $p$ distinct legal matchings and put them
into a list
$\cal L$; the largest energy of the matchings in this
list is
an upper bound $E_{UB}$ on the $p$th energy level for
our
system. Then we start at the
level of the tree's root and consider all of its
segments;
for each choice of segment, the search problem
corresponds to
finding matchings on a smaller system with one less
site
in each of the $d$ sets; the search can thus be
implemented
recursively. Suppose we have done $k$ recursions; the
sub-problem is associated to the node on our tree that
is
obtained by following the choices of hyperedges in the
recursive construction. This node
corresponds to a partial matching in which
the first $k$ sites of the first set have each been assigned
a hyperedge. An important property is that
all hyperedges have positive energies; then
we know that any matching that is compatible with
the current partial matching has an energy greater
than it,
thereby providing a lower bound on all the leaf
energies
obtainable from the current node. If that lower bound
is greater than $E_{UB}$, then the subtree rooted on the
current node can be pruned (discarded from the
search); otherwise, one iterates the recursion (that is one performs
branching on the different choices of the hyperedge
to include at the present level) and
$k$ goes to $k+1$. When this process leads to
a leaf that corresponds to a legal matching,
we compute the energy $E$ of this matching. If
$E<E_{UB}$,
we insert that matching into our list $\cal L$ and
remove
its worst element so that it always has $p$ elements;
we also update $E_{UB}$ which by definition is the
largest energy of the matchings in $\cal L$;
on the contrary, if $E>E_{UB}$, we discard the
matching (leaf).
After a finite number of branchings and prunings,
the algorithm has explored all choices for the
segments emerging from the tree's root and one is
done. The
best $p$ matchings are then in the list $\cal L$.

The algorithm without pruning requires $O(M^{M(d-1)})$
operations; with pruning and the different optimizations
sketched in Appendix~B, the number of operations
grows roughly by a constant factor when $M$ is
increased by $1$; in particular, for the random instances
studied here and $d=3$, this factor is about $2.2$.

\subsection{Ground state energies}
We generated a large number of random samples (disorder
instances with the hyperedge costs taken to be independent
uniformly distributed random variables in $\left[0,1\right]$)
and for each sample determined its ground state. We 
used several random number
generators to check that our results were robust. Because of the
exponential growth of the computation time with $M$, 
in practice we were limited to relatively modest values 
of $M$. For the results presented here 
and involving only ground states, at $d=3$
we used $10000$ samples for $M=20$ and $M=22$, while
for the smaller values of $M$ we used $20000$ samples.
We also performed runs at $d=4$ but with lower statistics
because the algorithm becomes less efficient as $d$ increases;
in fact, we were limited to $M \le 14$ for that case and had 
only $5000$ samples for each $M$.

Let's first focus on the behavior of ground-state energy. For 
each sample, we determine with our Branch \& Bound algorithm
the ground-state energy density $e_0 \equiv E_0/M \equiv M^{d-2} C_M^{(d)}$
[cf. Eq.~(\ref{eq:def})]; then we can analyse its mean in our ensemble
or consider other properties of its distribution.

\begin{figure}
\begin{minipage}[t]{.46\linewidth}
\centering
\epsfig{file=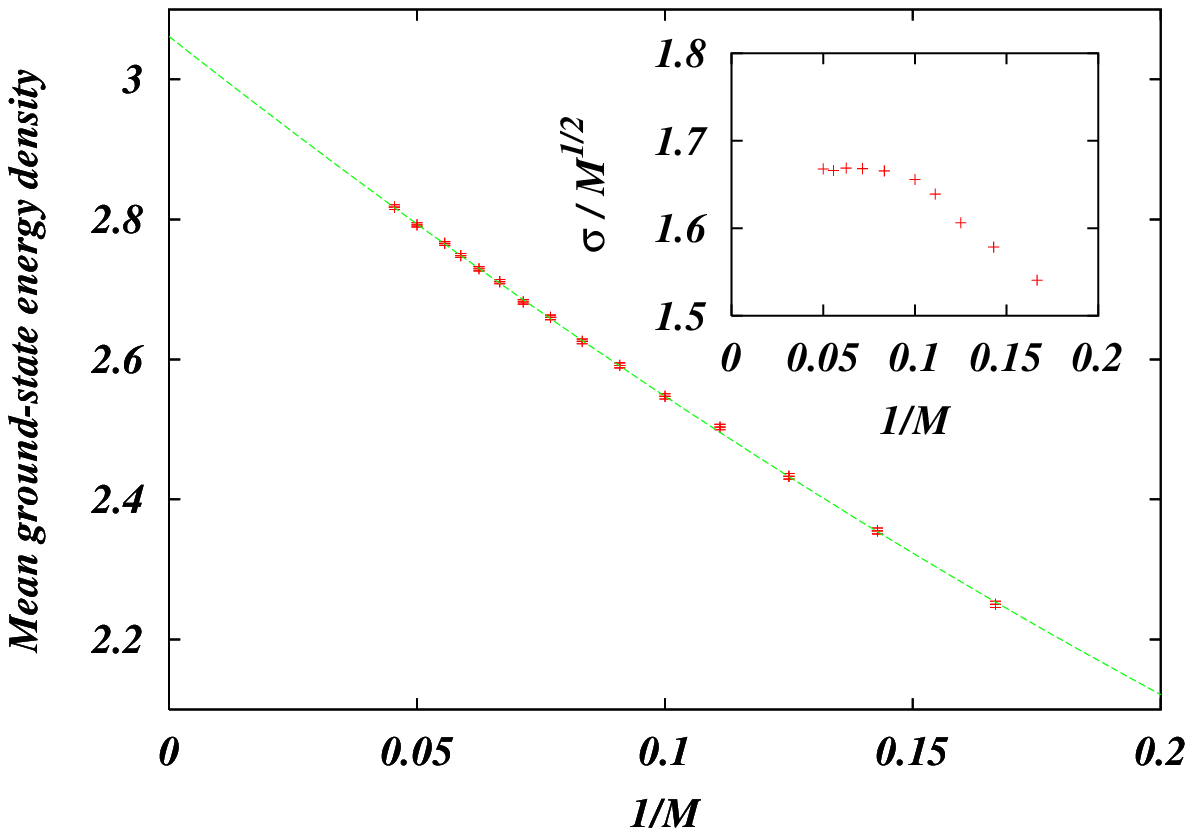,width=.9\textwidth}
\caption{\small Mean ground-state energy density
as a function of $1/M$ at $d=3$. The line is
the quadratic fit using $M\ge 10$ data.
Inset: the rescaled standard deviation of the
ground-state energy, suggesting
a central limit theorem behavior.
\label{fig:E_0}}
\end{minipage} \hfill
\begin{minipage}[t]{.46\linewidth}
\centering
\epsfig{file=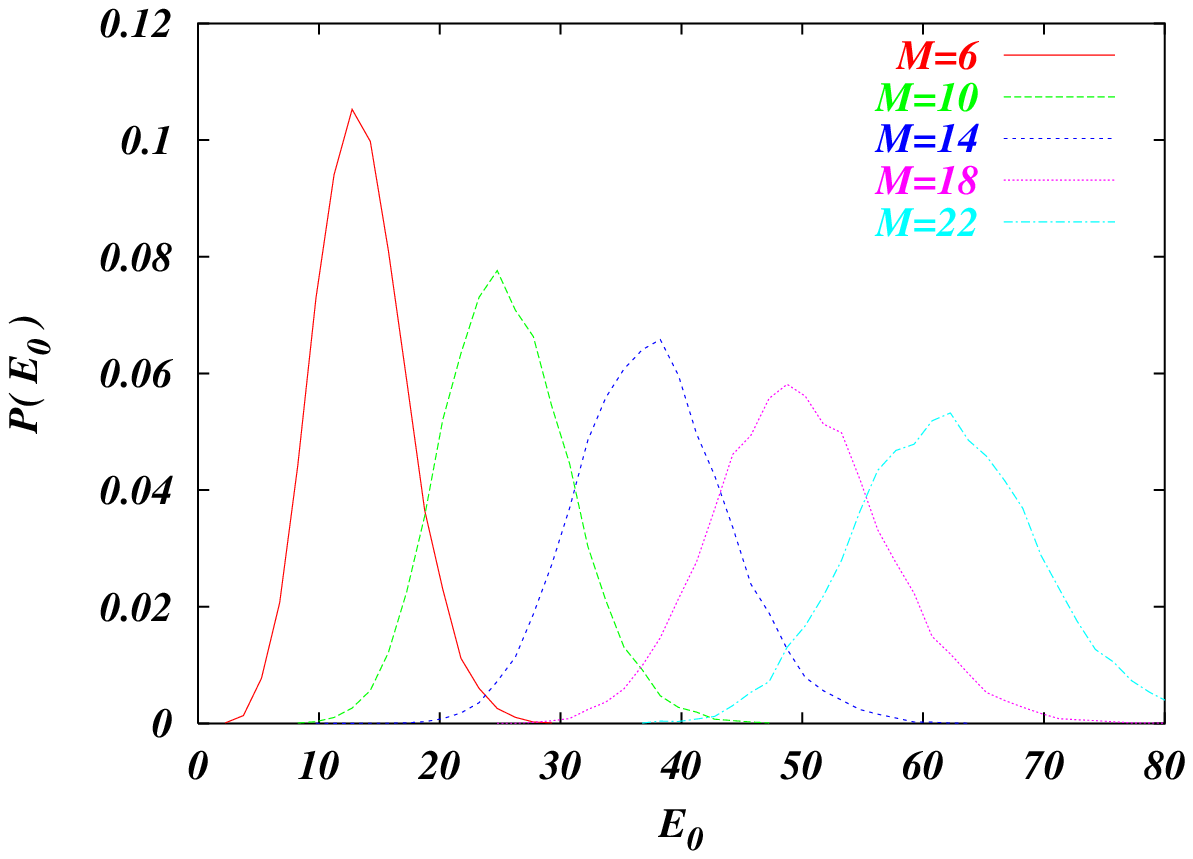,width=.9\textwidth}
\caption{\small Distribution of the extensive
ground-state energy for increasing $M$ values
(from left to right) at $d=3$.\label{fig:distE_0}}
\end{minipage}
\end{figure}
In Fig.~\ref{fig:E_0} we show how the mean ground-state 
energy density $\E[ e_0 ]$ changes as one increases 
$M$. The behavior is roughly linear
in $1/M$, but by eye one can definitely see some curvature.
Because of this, linear fits do not give good values 
of $\chi^2$ unless the $M < 10$ data are ignored; 
for instance, keeping only the $M \ge 10$ data, the linear
fit gives $3.040(3)$ as the limiting value
with $\chi^2=3.6$ for $9$ degrees of freedom, while
if we use all the data we obtain
$3.021(3)$ with $\chi^2=32$ for $14$ degrees of freedom.
We have also tried corrections of the type
$\ln(M)/M$ but this did not work well. Thus
we proceed by considering 
quadratic fits. In that case, the resulting $M=\infty$ intercept
does not depend much on whether one uses all or just
the highest values of $M$. In particular, for all the data,
we get the limiting value 
$3.046(5)$ with $\chi^2=9.6$ for $13$ degrees of freedom,
while using the $M \ge 10$ data only one has
$3.06(1)$ with $\chi^2=2.3$ for $8$ degrees of freedom.
(In all these estimates, the error bars quoted are statistical
only, as obtained from the statistical fluctuations.)
We have also considered power fits, namely
$\E[ e_0 ] = a + b/M^c$. Fitting all the data gives
the limiting value $3.08(1)$ with $\chi^2=7.2$ for 
$13$ degrees of freedom while keeping only 
the $M \ge 10$ data leads to 
$3.09(3)$ with $\chi^2=2.3$ for $8$ degrees of freedom
(in both cases, the exponent $c$ is close to $0.88$).
Since these $\chi^2$ are similar
to those of the quadratic fits, we see that the
systematic errors are not negligible and are at least
of the same order as the statistical errors; because
of these effects, the agreement with 
the theoretical value of $3.126$ can be considered rather good.

We studied similarly the case $d=4$. The data again has
positive curvature when plotted as a function of $1/M$,
but since we have less statistics and a much smaller
range of $M$, much less precision can be obtained
for the large $M$ limit. For the linear fit ($M \ge 9$)
we get a limiting value of $6.75(3)$ with $\chi^2=4.7$ 
for $4$ degrees of freedom. For the quadratic fit
($M \ge 9$ again), we get $7.22(8)$ 
with $\chi^2=0.37$ for $3$ degrees of freedom.
Finally, for the power fit we get $10.2(9)$ 
with $\chi^2=1.0$ for $5$ degrees of freedom; the exponent
is $c=0.3$ which is small and leads to a
large upturn for $M > 100$; clearly that regime is far beyond our
reach and suggests that the power fit is probably 
inappropriate as non robust (note for instance that
the uncertainty on the limiting value is far higher
here than for the other fits). The different estimates
show that uncertainties arising from systematic effects
($M$ too small) are severe; instead of the $1\%$ precision
we had at $d=3$, we have a precision of at best $10\%$ at
$d=4$ (compare to the theoretical prediction of $7.703$).
The conclusion is that numerics do not teach us much
for the case $d=4$ and so hereafter we shall concentrate
on the different properties arising when $d=3$.

One of the expectations for the $d$-index matching problem
is that the free energy is self-averaging. Although
at present there is no proof of such a property, there is no reason
to expect otherwise; here we are limited by the
numerical approach to
ground states, but in that framework we can determine empirically
the \emph{distribution} of energies in the ensemble of random 
instances. Fig.~\ref{fig:distE_0} displays the probability
distribution of the (extensive) ground-state energy $E_0$
for several values of $M$ ($d=3$). If as expected, the 
ground-state energy is self-averaging, the relative width
of these distributions should go to zero. We have thus
measured the first few moments of these distributions.
In the inset of Fig.~\ref{fig:E_0}, we have plotted
the standard deviation $\sigma$ of the ground-state
energy divided by $\sqrt{M}$ as
a function of $1/M$. Self-averaging
corresponds to having $\sigma/M \to 0$; from the inset
we see that $\sigma/M^{1/2}$ goes to a constant 
at large $M$ so self-averaging holds and the convergence of
the distribution is compatible with
a central limit theorem type behavior; such a
scaling arises from sums of not too dependent
random variables and leads to a Gaussian limiting
shape. To confirm this, we have looked at higher moments:
we find that indeed the skewness and kurtosis of the
distributions decrease, in line with a central limit
theorem type convergence. 

Having a limiting Gaussian distribution for $E_0$ is not a consequence
of the frozen 1RSB pattern of replica symmetry breaking
since in the random energy model the distribution
of $E_0$ follows a Gumbel distribution; furthermore, in that
case the fluctuations in $E_0$ are $O(1)$ whereas in the
matching problem they are $O(\sqrt{M})$. To see why such
large fluctuations are ``natural'', consider instead
of $E_0$ the quantity $\mathcal{E}_0$ obtained by
adding the lengths $\ell_i$ of the shortest hyperedges containing
each site $i$ of the first set. This quantity
arises in a greedy algorithm (but
which does not necessarily generate a legal matching)
and clearly one has $E_0 \le \mathcal{E}_0$.
The central limit theorem applies to $\mathcal{E}_0$, so it
will have a standard deviation that grows as $\sqrt{M}$
and its distribution will become Gaussian at large $M$.
The actual ground-state energy $E_0$ is obtained by
allowing hyperedge lengths that are slightly larger than the
$\ell_i$, but this should not suppress the
large fluctuations nor prevent the central limit theorem scaling.

\subsection{Other ground state properties}

\begin{figure}
\begin{minipage}[t]{.46\linewidth}
\centering
\epsfig{file=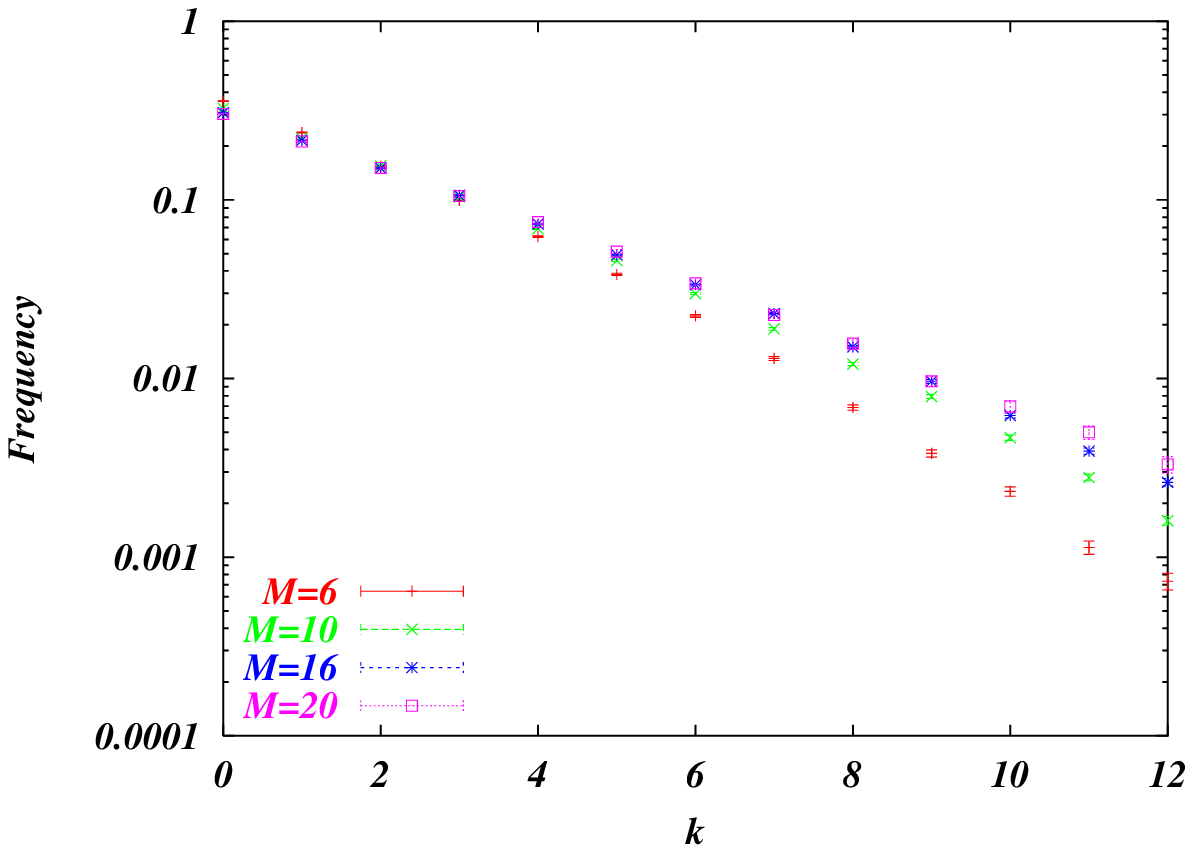,width=.9\linewidth}
\caption{\small Histogram of the occupation probabilities
in the ground state of the hyperedges as a function
of their order $k$ ($d=3$). (Order is $1$ for the lowest
value among those hyperedges containing a given site, 2 for the
next lowest value etc...) At large $k$,
these frequencies approach an exponential law.\label{fig:orders}
}
\end{minipage} \hfill
\begin{minipage}[t]{.46\linewidth}
\centering
\epsfig{file=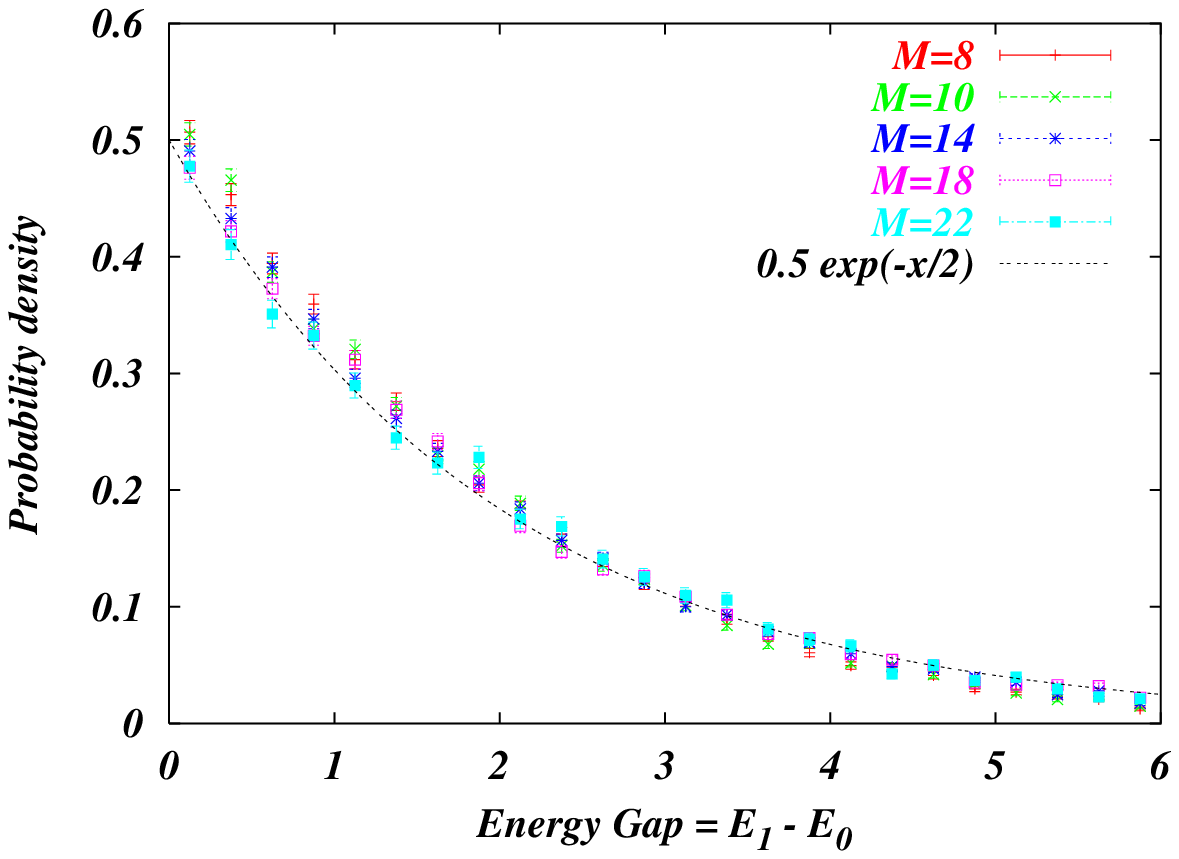,width=.9\textwidth}
\caption{\small Probability density of the gap,
$E_1 - E_0$, that is the energy difference between the
first excited state and the ground state (extensive) energies in the case
$d=3$. The curve is a pure exponential to guide the eye.\label{fig:PofGap}}
\end{minipage}
\end{figure}

As discussed at the beginning of this paper, one
expects the hyperedge containing
a given site in the ground state matching to be one
of the shortest possible ones. To investigate this
issue quantitatively, let us order
all the hyperedges containing a given site, going
from the shortest to the longest hyperedge. The
``order'' of a hyperedge is then $1$ if it is the shortest, $2$ 
if it is the next shortest, etc... The orders arising in the
ground state should be dominated by the lowest ones, 
$1$, $2$, $3$... Consider thus the frequencies 
with which these orders arise; in Fig.\ref{fig:orders} we 
show the behavior of these frequencies for increasing $M$ in 
the case $d=3$. We see that there is a limiting histogram at large
$M$, and that indeed the lowest orders dominate. Furthermore,
we see that for large $k$ the probability of
occupation of an edge tends to decrease exponentially with $k$
(the data are displayed on a semi log plot).
Note that in the standard matching ($d=2$) problem, the decrease 
goes as $1/2^k$ exactly, while for our $d=3$ case, the exponential
decay is only asymptotic; furthermore, we have found no simple
expression giving the decay rate of this exponential.

\subsection{Excited states}
Let us consider now states above the ground state.
Define the excitation energy or ``gap'' as
$E_1-E_0$ where $E_0$ is the extensive ground-state energy
and $E_1$ that of the next lowest energy state. In
Fig.~\ref{fig:PofGap} we show that this random variable has a limiting
distribution so that $E_1-E_0 = O(1)$ in the
large $M$ limit, just as happens in the random energy model. Furthermore,
the distribution is very well fit by an exponential (cf. the curve shown
in the figure).

\begin{figure}
\begin{minipage}[t]{.46\linewidth}
\centering
\epsfig{file=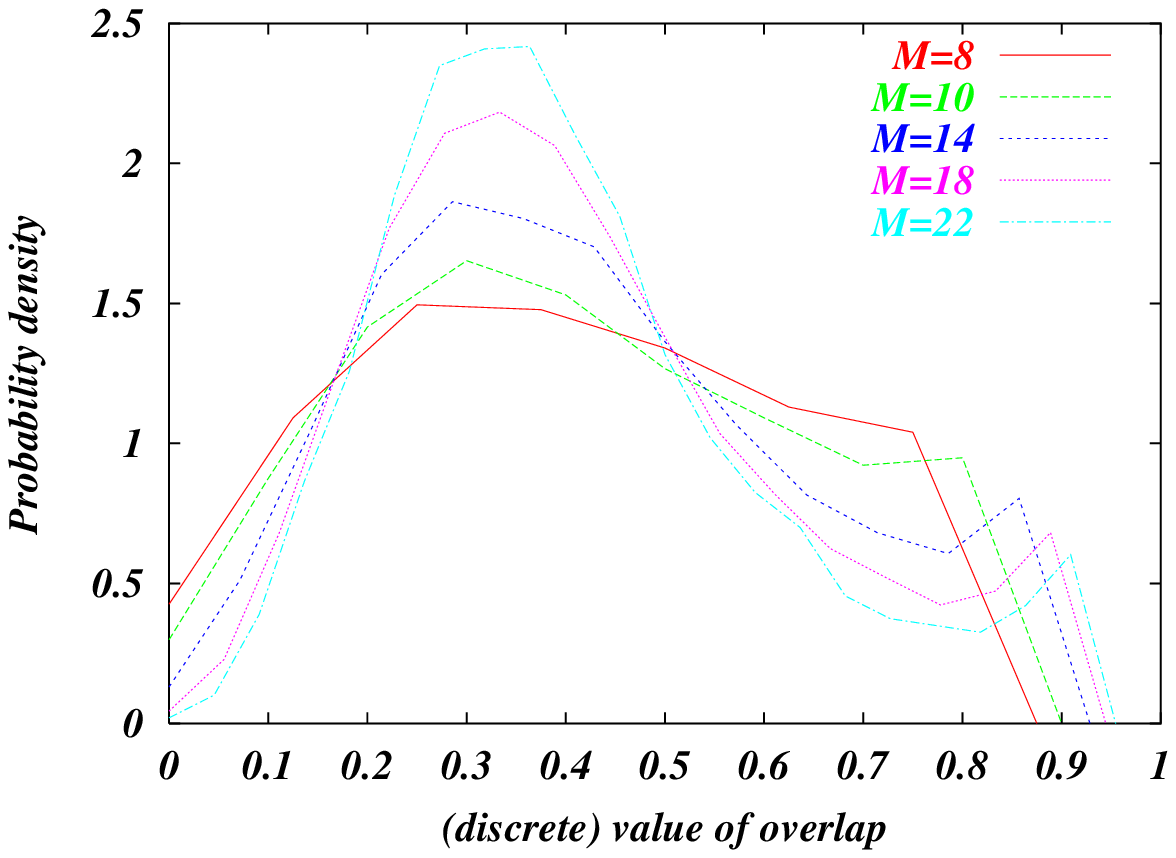,width=.9\linewidth}
\caption{\small Probability density of the overlap $q$
between the ground state and the first excited
state for increasing $M$ ($d=3$).\label{fig:PofQ}}
\end{minipage} \hfill
\begin{minipage}[t]{.46\linewidth}
\centering
\epsfig{file=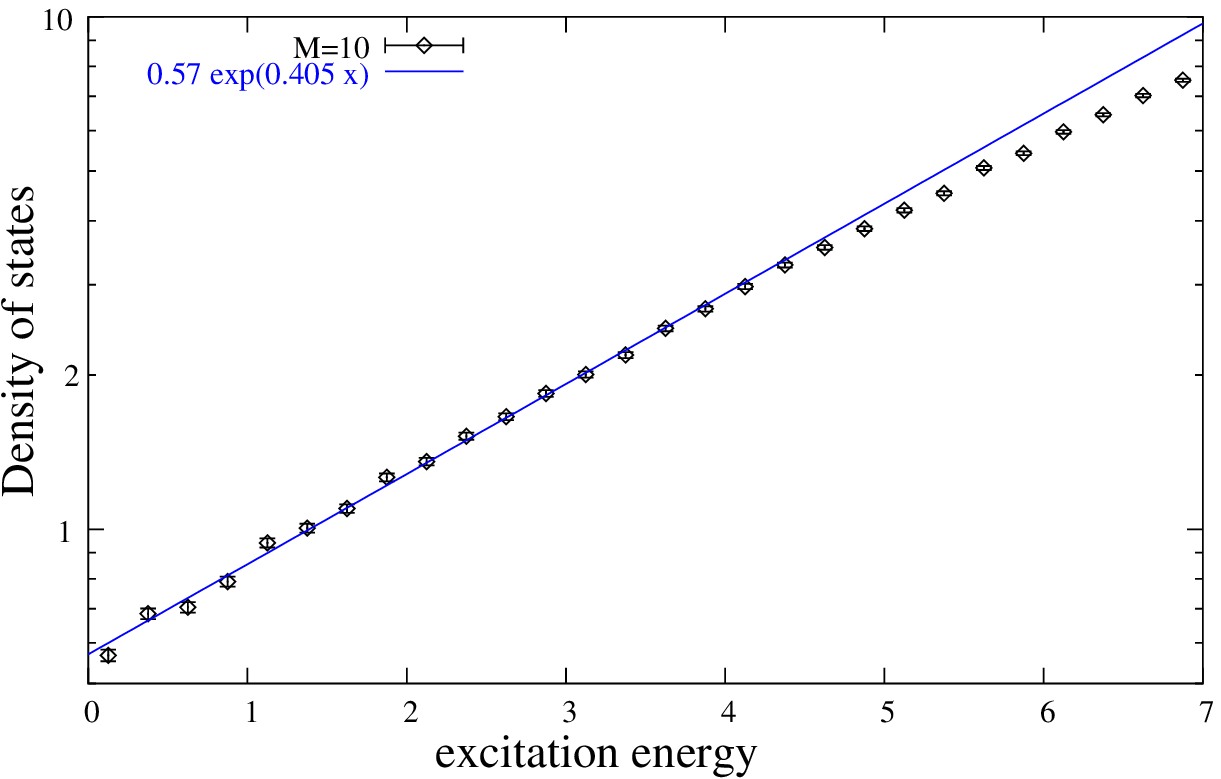,width=.9\textwidth}
\caption{\small Density of energy levels, measured
from the ground-state
energy. At low energies (before
finite size effects dominate), this density grows
exponentially
as $\exp[\beta_c (E-E_0)]$ thereby giving the model's
critical temperature. Shown is the case
$d=3$ for $M=10$.\label{fig:density}}
\end{minipage}
\end{figure}

Following our theoretical conclusions obtained earlier,
consider now the overlap between the ground state and
the first excited state. In our frozen 1RSB picture, these
matchings are expected to have a fixed
(self-averaging) overlap when $M$ grows. In Fig.~\ref{fig:PofQ} we
show the probability distribution of such overlaps for
increasing $M$. We see that there is a local peak at large
overlap that shifts toward $q=1$ but which simultaneously
decays. The bulk of the overlaps however arise around $q=0.3$ and
when $M$ increases we see that the corresponding peak both
gets higher and more narrow. Overall, the behavior
is compatible with a convergence toward a Dirac peak
near $q=0.32$, to be compared with the theoretical
prediction $q_c = 0.321$.

\subsection{Low energy entropy}

Finally, consider the \emph{density} of energy levels. In the case of
the random energy model,
this density becomes self-averaging when
the excitation energy grows. We have thus computed the
disorder averaged density of levels as a function of the
excitation energy, $E-E_0$. That is a measure of the
exponential of the microcanonical entropy; within
the frozen 1RSB scenario, it gives the critical temperature via
$\rho(E-E_0) \sim \exp[ \beta_c (E-E_0)]$.
In Fig.~\ref{fig:density} we display our numerical
estimate of $\rho$ and see that it is very nearly a pure
exponential. From the slope on the semi-log plot we
extract $\beta_c \approx 0.405$; this value should be
compared to the theoretical prediction of $0.412$; the agreement
is reasonable but not perfect. To get
better agreement, we believe it would be necessary to
go to larger $M$ and also to go further in the
self-averaging regime, i.e., to consider larger
$E-E_0$ which numerically is an arduous task.

\section{Conclusion}

We presented an analysis of multi-index matching problems (MIMPs) based on an adaptation of the cavity method for finite connectivity systems. For the well-known two-index matching problem, our approach provides an alternative derivation of results previously obtained using the replica and cavity methods. With respect to these older studies, the present one has the advantages of being closer to the mathematical framework developed by Aldous, and of allowing replica symmetry breaking effects to be incorporated in a tractable manner. Exploiting this latter possibility, we predict the value of the asymptotic minimal cost to be given for $d$-index matching problems by $\mathcal{L}^{(d)}=\epsilon_{RS}(\beta_s)$ with $\epsilon_{RS}$ obtained from Eq.~(\ref{eq:Grec}) and (\ref{eq:eands}) and $\beta_s$ satisfying $s_{RS}(\beta_s)=0$. Formally, this $d\geq 3$ conjecture differs from the case $d=2$ (where it is a theorem) in that $\beta_s=\infty$ when $d=2$, while $\beta_s<\infty$ when $d\geq 3$. The distinction between 2-index and $d$-index matching problems with $d\geq 3$ arises clearly from our analytical and numerical analysis: in the first case all low cost matchings share most of their hyperedges, while in the second case they differ from each other by a finite fraction of their hyperedges. In mathematical terms, the essential uniqueness property does not hold when $d\geq 3$ or in physical terms replica symmetry must be broken. Extending Aldous' framework to rigorously account for this fact and providing a proof of our conjecture for $d\geq 3$ seems to us a particularly interesting mathematical challenge.

From a physical perspective, the qualitative difference between 2-index and $d$-index matchings problems with $d\geq 3$ hinges on the presence at low temperature of a glassy phase. This is similar to the difference that has been found between the 2-SAT and 2-coloring problems, which are polynomial, and the $K$-SAT and $q$-coloring problems with $K\geq 3$ and $q\geq 3$, which are NP-complete. For MIMPs, the nature of the glassy phase is however simpler, as it is made of isolated configurations instead of separate clusters of many configurations. We termed this phase a ``frozen 1RSB glassy phase'' and attributed it to the nature of the constraints, called hard constraints. As a technical consequence of this distinctive feature, a particular frozen 1RSB Ansatz has to be implemented. Such an Ansatz has repetitively been used in the literature as a convenient (but rarely justified) substitute for the more complicated general 1RSB Ansatz; our discussion on the role of hard constraints provides a clarification of its conditions of validity which we believe is of general interest for the investigation of glassy phases in other systems.

\begin{acknowledgments}

This work was supported in part by the European Community's
Human Potential Programme under contracts 
HPRN-CT-2002-00307 (DYGLAGEMEM) and
HPRN-CT-2002-00319 (STIPCO) as well as by the Community's
EVERGROW Integrated Project.

\end{acknowledgments}

\appendix

\section{Population dynamics algorithm}

Here we give a short description of the population
dynamics algorithm we used to solve the RS cavity equations. We
implemented two different versions, corresponding to the two
different cut-off procedures mentioned in the text, associated either
with Poissonian (algorithm P) or regular graphs (algorithm R). In 
addition to the inputs $d$ and $\beta$, the algorithm has essentially 3
parameters: the mean degree of
the nodes, $C$ (algorithm P) or $K$ (algorithm R), the size of the population,
$\N_{\rm pop}$, and the number of iterations $\N_{\rm iter}$. The 
common structure of the two algorithms is the following:
\begin{itemize}
    \item Initialize with random values a population of cavity fields $x[i]$,
    $i=1,\dots,\N_{\rm pop}$;
    \item Do $\N_{\rm trans}=100$ times: $Update()$;
    \item Do $\N_{\rm iter}$ times: $Update()$ and $Measure()$.
\end{itemize}
The first loop allows the system to equilibrate toward
the stationary distribution. The subroutine $Update()$ depends
on the cut-off procedure and can schematically described as
follows:\\

Do $\N_{\rm pop}$ times:
\begin{itemize}
    \item Draw $k$ either at random with Poissonian
distribution of
    mean $C$ (algorithm P), or take $k=K$ (algorithm R);
    \item Draw costs $\{\xi_a\}_{a=1,\dots,k}$ either
independently with the uniform
    distribution in $[0,C]$ (algorithm P), or according to a Poisson
process with rate 1 (algorithm R);
    \item Draw at random $k(d-1)$ members of the
population and use them together with the $\xi_a$ to
    compute a new field $x_0$ according to Eq.~(\ref{eq:hatx});
    \item Draw at random one member of the population
and replace its cavity field value with $x_0$.
\end{itemize}

The subroutine $Measure()$ is implemented similarly and 
computes the free energy according to 
Eqs.~(\ref{eq:fRS1})-(\ref{eq:fRS2}). The final
output for the
free energy is obtained by averaging over the $\N_{\rm
iter}$
iterations, while the fluctuations across iterations
are used to
check convergence. The algorithm must be run for
increasing values
of $C$ or $K$ to extrapolate the $C\to\infty$ (algorithm P) or
$K\to\infty$ (algorithm R)
limit, requiring one to consider larger and larger population sizes
$\N_{\rm pop}$
to obtain reliable results. Taking this limit is
however
facilitated by the numerical observation that the
Poissonian
approximation (algorithm P) approaches the solution from below while the regular approximation (algorithm R)
approaches
it from above; this is illustrated in Fig.~\ref{fig:f4drs} 
with $d=4$. We refer to the captions of the various 
figures for typical choices of the parameters $C$, $\N_{\rm pop}$ 
and $\N_{\rm iter}$. The numerical results we obtained 
for $d=2$ are consistent with the exact solution, $\beta_c=\infty$ 
and $f_{RS}(\beta_c)=\pi^2/12$, and are the following for $d=3,4$:
\begin{equation}
\begin{split}
d=3:\quad \beta_c=0.412\pm 0.001,\quad f_{RS}(\beta_c)=1.042\pm 0.0003,\\
d=4:\quad \beta_c=0.135\pm 0.002,\quad f_{RS}(\beta_c)=1.925\pm 0.0006.
\end{split}
\end{equation}
The free-energy densities are given here for simple matching 
problems and their counterpart for $d$-partite matchings are 
obtained by multiplying the values by $d$.

We have also implemented the generalization of this
algorithm to solve the 1RSB cavity equations~(\ref{eq:1rsb}) and used
it to check that no discontinuous transition occurs prior to
the entropy crisis (see~\cite{MezardParisi01} for algorithmic details).

\section{Aspects of the branch and bound algorithm}

Our objective is to solve $d$-partite matching problems
at sufficiently large $M$ so that an extrapolation to the
$M \to \infty$ limit can be performed without too much
uncertainty. For many problems (satisfiability,
coloring, etc...), one prefers an easily implementable algorithm
such as one in the class of ``heuristic'' algorithms; in such approaches
one performs a fast search for the ground
state but no guarantee is provided that
the global optimum will be found. Examples of these algorithms are
simulated annealing and variable depth local search.
Heuristic algorithms typically attempt to move towards
regions of lower energy by searching in 
the neighborhood of a current configuration. However,
since the search is local, such an approach is bound to break down for problems
in which the frozen 1RSB scenario applies. This fact pushed us towards
the development of an ``exact'' algorithm capable of delivering a 
certificate of optimality of the proposed ground state.
Amongst exact algorithms, enumeration can be discarded
because it is much too slow; Branch \& Bound gets around
this problem through pruning of the enumeration/search.
There are also other possible methods such as Branch \&
Cut, but these require an in depth understanding
of polytopes and rely on separation procedures
which have not yet been developped for MIMP. Note that in all
exact methods, the key to efficiency is to have good bounds; fortunately
MIMPs are relatively well adapted to such a 
strategy.

We already discussed in the main text our choice of representation
of matchings and partial matchings. Given a partial matching
of the first $k$ sites of the first set, we have to solve a
MIMP with $M-k$ sites and so the algorithm can be implemented 
recursively. Since at each node we need to consider all of its
possible branchings (naively, there are $(M-k)^{d-1}$ of these),
it is useful to order these branchings according to the
length of the corresponding hyperedges, going from short to 
long. Rather than recompute these orderings dynamically every time
the partial matching changes, we do it once and for all
at the initialization of the program. This allows for speed
but it must be compensated by a rapid determination of whether
a given hyperedge is allowed; for that we use a data structure
which tells us for each site of each set whether it is
matched (belongs to one of the occupied hyperedges). 
This structure is updated whenever a partial matching
is extended or reduced.

The pruning of the search must be as stringent as possible, and this
depends on the quality of the bounds. Our simplest
bound $B_1$ is just the current partial matching's energy $E_k$:
if that energy is higher than $E_{UB}$ (the upper bound $E_{UB}$ 
as defined in section~\ref{subsec:bb}), then the whole sub-tree
below the current node can be pruned. A better bound is $B_2$, obtained
by adding to $E_k$ the sum over each remaining unmatched site of the first set
of the shortest hyperedge containing that site. This sum can
be precomputed and tabulated. A still better bound is $B_3$
obtained as $B_2$ but where now one takes for each site
the shortest hyperedge that is compatible with the current partial
matching. This bound cannot be predefined once and for all
and is slow to compute. Since we have found it to be useful
for pruning, we have optimized its determination by noticing
that it can be tabulated and modified incrementally: every time
the partial matching is extended (a hyperedge is added),
we perform the search for the compatible hyperedges of each
unmatched site starting from the index (order) previously 
found to be compatible. When backtracking, one has to remove
a hyperedge and there we simply go back to the tables
we had at that level: in effect, we maintain efficiency
if we assign tables at each level and follow their updating
one step at a time.

The rate of pruning is very different for the three bounds,
and we found that a good strategy (for balancing pruning
rate and computation time) was to apply the three bounds
successively: if the first one does not prune, one goes
on to the second one and so forth. To speed up the computation further, 
we found it useful to implement the recursivity of the program
in a limited mode only: the data structures are set up
once and for all at initialization time, the hyperedges
are ordered once and for all too, and then the recursion is used
mainly to go through the branchings and to maintain the
tables. Efficiency is gained as no reorganization of the instance (hyperedge
weights) is performed, and in particular no ``smaller matching
problem'' is ever defined explicitly.

\bibliographystyle{apsrev}

\bibliography{reference,matchings,glasses,graphs}

\begin{thebibliography}{39}
\bibitem[{fnV()}]{fnViolationAnnealed}
\bibinfo{note}{An additional evidence that the na{\"\i}ve generalization of the
  formulae used when $d=2$ fails for $d\geq 3$ is found when $d=4$ (extending
  presumably to all $d\geq 4$) where the RS prediction for
  $\mathcal{L}^{(d)}_{RS}$ violates the annealed bound.}
\end{thebibliography}

\end{document}